\documentclass[preprint,prb,showpacs]{revtex4}

\usepackage[dvips]{graphicx}
\usepackage[dvips]{color}

\usepackage{graphicx}
\usepackage{epsfig}

\newcommand{\be}{\begin{eqnarray}}
\newcommand{\ee}{\end{eqnarray}}
\newcommand{\la}{\langle}
\newcommand{\ra}{\rangle}

\begin{document}
\title{Equation-of-motion treatment of hyperfine interaction in a quantum dot}
\author{Changxue Deng}
\author{Xuedong Hu}
\affiliation{Department of Physics, University at Buffalo, SUNY,
Buffalo, NY 14260-1500}
%\date{\today}

\begin{abstract}
Isolated electron
spins in semiconductor nanostructures are promising qubit candidates for a
solid state quantum computer, 
There have seen truly impressive experimental
progresses in the study of single spins in the past two years. In this paper 
we analytically solve the {\it Non-Markovian} 
single electron spin dynamics due to inhomogeneous hyperfine couplings with 
surrounding nuclei in a quantum dot.  We use the equation-of-motion method 
assisted with a 
large field expansion in a full quantum mechanical treatment. We recover
the exact solution for fully polarized nuclei. 
By considering virtual nuclear spin flip-flops 
mediated by the electron, which generate fluctuations in the Overhauser field
(the nuclear field) for the electron spin,
we find that the decay amplitude of the transverse electron spin correlation 
function for partially polarized nuclear spin configurations is of the order
unity instead of $\text{O}(1/N)$ ($N$ being the number of nuclei in the dot) 
obtained in previous studies. 
We show that the complete amplitude decay can be 
understood with the spectrum broadening of the correlation function near the 
electron spin Rabi frequency induced by nuclear spin flip-flops. 
Our results show that a 90\% nuclear polarization can enhance the electron 
spin $T_2$ time by more than one order of magnitude in some parameter regime.  
In the long time limit, the envelope of the transverse electron spin 
correlation function has a non-exponential $1/t^2$ decay in the presence of 
both polarized and unpolarized nuclei.         
\end{abstract}
\maketitle

\section{Introduction}
The electron spin dynamics in semiconductor nanostructures is
presently of particular interest both experimentally 
\cite{Fuji_Nature_02, Koppens_Science_05, Petta_Science_05, Koppens_Nature_06}
and theoretically because of 
the potential applications in spin quantum computation. \cite{Hu_Rev}
Among the many roadblocks to solid state quantum computing, 
maintaining the quantum coherence of the electron spin 
in a quantum dot is a crucial issue.  One of the most important 
and relevant spin interactions for this confined electron is 
hyperfine coupling with the surrounding nuclei, or the so-called Fermi 
contact interaction. The role of hyperfine 
interaction could be both negative and positive depending on the 
actual experimental
operations. On the one hand, nuclear spin bath could be a major 
decoherence channel
for the electron spin that acts as a qubit. 
\cite{Loss_PRA_98, Taylor_Nature_05}
This problem is unavoidable in a GaAs quantum dot where 
all the isotopes of Ga and As has nuclear spin $I=\frac{3}{2}$, 
Although this is not a problem for silicon quantum
dot where the nuclei of $^{29}$Si ($I=\frac{1}{2}$, with natural 
abundance 4.68\%) could be removed, so that the silicon is made up of 
only $^{28}$Si or $^{30}$Si, both having 
nuclear spin $I=0$. On the other hand, 
electron-nuclei hyperfine interaction could also be utilized to 
control the nuclear spins, which as an ensemble could act coherently. 
For example it has recently been suggested that the 
ensemble of nuclear spins in a quantum dot may be 
used as a long-lived quantum memory for electron spin by 
transferring the electron spin dynamics to nuclear spin 
reservoir coherently.\cite{Taylor_03} 

%To study the relaxation and decoherence of electron spin in the dot 
%two different approaches of treating the hyperfine interaction
%exist in the literatures. 
%Roger and Das Sarma \cite{Roger1,Roger2} have taken the $z$ component 
%of hyperfine coupling (the first term in $H_h$)
%as an inhomogeneous magnetic field (the Overhauser field) which 
%is felt by electrons in 
%the dots. They calculate the nuclear spin flip-flop rates due to 
%the inhomogeneity
%of hyperfine coupling constants and the resulting electron spin decoherence. 
%A exponential decay has been found using Markovian approximation 
%for nuclear spins.
%Later similar line has been applied to ca calculate the suppression of 
%nuclear spin diffusion. 
%\cite{Deng_sd} In these calculations, the electron-nuclei spin 
%exchange (flip-flop 
%between electron and nucleus) has been neglected by assuming there 
%is an non-vanishing
%external magnetic field which leads to very large Zeeman energy 
%mismatch between electron 
%and nuclei. In another approach, the full hyperfine Hamiltonian 
%(see Eq. \ref{ham_spin} below) 
%has been treated consistently by ignoring the dipolar interaction 
%among nuclear spins 
%which is a good approximation because the dipole-dipole interaction 
%is several orders smaller
%than the hyperfine coupling. The decoherence and relaxation of 
%electron spin are also
%caused by the inhomogeneous hyperfine coupling with the nuclei 
%located at various sites
%within the quantum dot.

The study of Non-Markovian electron spin dynamics in the 
presence of hyperfine interaction is a complicated problem 
due to its quantum many-body (electron and nuclei) nature. The 
problem has been studied by many researchers previously. 
\cite{Khaet_03,Schli_PRB_02,Erlin_PRB_04,Coish_PRB_04, Shenvi_PRB_05,
Yao_2005, Deng_PRB_06} 
An exact solution \cite{Khaet_03} has been found 
for an electron interacting with the fully polarized nuclear reservoir.  
Various types of approximations, perturbation 
theory \cite{Coish_PRB_04} or an effective Hamiltonian approach, 
\cite{Shenvi_PRB_05, Yao_2005} 
have been employed in dealing with this problem. 
Numerical studies, \cite{Schli_PRB_02,Erlin_PRB_04,Shenvi_PRB_05} 
limited by the exponentially large Hilbert space,
can only be applied to small systems, typically of up to
20 spins. Here we analyze the problem starting from the exact 
Hamiltonian using the equation-of-motion (EOM) method by working in 
the Heisenberg picture. Usually the problem is studied using the master
equations for the system density operator. \cite{Book_Weiss}
We develop a systematic large field 
expansion method to simplify the equations dramatically. This 
way we can address the full quantum mechanical problem 
analytically using well-controlled approximations.

A confined electron in a gated quantum dot generally 
has a nonuniform envelope wavefunction and a relatively large radius,
which leads to an inhomogeneous hyperfine 
coupling with the nuclear spins on the crystal lattice. 
The effective number $N$ of nuclear spins involved is $10^4-10^6$ 
depending on the actual size of the quantum dot. It is exactly 
this non-uniformity that causes both electron spin relaxation 
and dephasing. Physically this is because the electron acquires 
a different phase factor each time it interacts with a particular 
nuclear spin with random orientation. This effect is totally 
quantum mechanical, and is quite 
different from the dephasing effect caused by inhomogeneous broadening 
or ensemble average ($T_2^*$), \cite{Merkulov_PRB_02}
which can be removed by spin echo technique. 
\cite{Koppens_Science_05, Petta_Science_05}

Electron mediated nuclear spin interactions have been 
studied for a long time, in both metals and in semiconductors.
\cite{Book_NMR_S, Bloem_PR_55}
%In this paper we focus on the problem of the spin decoherence
%of a single electron in the QD due to hyperfine interaction with the 
%surrounding nuclear spins. 
Although at a finite magnetic 
field the direct electron-nuclear-spin flip-flop is highly 
unlikely due to the Zeeman energy mismatch, higher-order 
processes where electron spin does not flip are possible. 
For example, conduction-electron-mediated nuclear spin 
interaction [Ruderman-Kitterl-Kasuya-Yoshida (RKKY)] has 
been studied for a long time in both metals and 
semiconductors. Here we focus on the electron-mediated 
RKKY interaction between nuclear spins on {\it single}
mediating electron spin. 
Specifically we focus on the backaction of this effective nuclear
spin interaction on the mediating electron spin and 
consequent electron spin decoherence.
Helped by a systematic large 
field expansion, we solve the full quantum mechanical problem
analytically and reveal the crucial importance of the 
electron-mediated nuclear spin flip-flop processes in the 
decoherence of an electron spin.

We note that virtual (high order) processes can be classified 
into two categories, one involving electron spin flip and the 
other without electron spin flip. In a high effective field 
(external plus nuclear field), these two types of processes 
have completely different consequences.  In terms of energy 
the process with electron spin flip needs large extra energy 
to compensate the electron Zeeman energy, since the nuclear 
Zeeman energy is much smaller and can be neglected.  Such 
processes (real processes for the electron), no matter
whether they  
first order or higher orders, are highly unlikely for 
large fields. Not surprisingly a negligible decay amplitude 
of order $O(\frac{1}{N})$ for both diagonal (relaxation) 
and non-diagonal (dephasing) components of the density matrix 
element of 
electron spin was found from these processes. In the second 
category, no extra energy is necessary since there is no electron 
spin flip between the initial and final states. 
It is found these kinds of processes indeed 
have significant contributions to $T_2$ using numerical simulation, 
\cite{Shenvi_PRB_05}
though for only 13 nuclear spins from an effective Hamiltonian.  To 
illustrate the role of virtual flip-flopping of nuclear pairs, 
we shall compare the exact solution of fully 
polarized nuclei where no virtual nuclear spin-exchange is 
allowed with that of the case in which one of the nuclear 
spins is in the down state initially, assuming that all the
nuclear spins point up in the fully polarized configuration.

%Along the second line mentioned above, the problem has been dealt by a lot of authors 
%using either the usual perturbation theory up to the fourth order beyond the Born approximation, 
%semi-classical approximation and exact numerical
%diagonizations for a not fully-polarized nuclear spin configuration. It is found that
%the longitudinal component of electron spin 
%$\la S^z(t)\ra$ generally has a non-exponential decay. However the perturbation calculations 
%can not be trusted in the case of zero external field and vanishing nuclear spin polarization
%where there is no small parameter that can be treated as a perturbation; exact numerical 
%diagnization can be only applied to nineteen nuclear spins due to the huge dimension of the
%Hilbert space. 

%In the present paper we attack this problem using a large field expansion 
%technique combined with 
%the equation-of-motion method in a full quantum mechanically treatment. 
%The expansion parameter
%$\frac{N}{4\Omega}$ here is the effective number of nuclear 
%spins within a quantum dot. Since $N$ is a very large number ($10^4-10^6$, 
%depending on the 
%actual sizes of the dots), our approximation should be 
%very accurate in principle. On the other
%hand, our method are also robust in dealing with non-perturbative limit 
%where there is no external 
%magnetic field and nuclear spin polarization. Our large N expansion 
%is essentially different from
%those which has been applied in solving the Kondo or impurity problem 
%where the parameter $N$ is
%the orbital angular momentum degeneracy of the $f$ electrons or 
%the localized states.\cite{LargeN}

In general, exact solution cannot be found for a 
non-quadratic Hamiltonian. From the perspective of
equations of motion, this means that there is 
an infinite number of equations when one attempts to 
evaluate the Green's function. Therefore, a decoupling 
method has to applied to cut off the series of equations 
and close the system. In the usual many-body treatment, this 
amounts to calculate some {\it thermal averages} using the
spectral functions determined from the Green's functions 
self-consistently. However, such approximation 
involving ensemble average
is not available in our present case, where we are interested in 
the {\it real time} dynamics including the coherent part and the 
decaying part of a single quantum many-body 
system. In our study we assume that the effective magnetic 
field (denoting the total of external and  nuclear fields) 
is large so that the electron Zeeman energy is much larger 
than the single nuclear spin Zeeman energy, and 
the direct electron-nuclei 
flip-flop is not allowed energetically. This assumption 
greatly simplifies our discussions by selecting one group of
equations, which give the major contribution to decoherence.
Typically, an effective field slightly larger
than the fully polarized nuclear field is enough for 
obtaining self-consistent result. Our study has 
experimental relevance since the full Overhauser field is
about 4 Tesla in GaAs quantum dots.

%Our large N expansion succeeds for the following reason. One first notice 
%that the time scale of
%the decoherence of electron spin is much slower that of the coherent 
%oscillations. For instance,
%it has been understood that the time scale for the onset of the decay is 
%of order 1, while 
%the period of oscillations $\sim$ $\sqrt{N}$ for unpolarized case and 
%$\sim N$ for the highly 
%polarized case. Now when calculating the time derivative of electron 
%spin components, the contribution
%of the amplitude decay is of order $\frac{1}{N}$ or $\frac{1}{\sqrt{N}}$ 
%smaller due to the
%coherent oscillation term ($e^{iNt}$ or $e^{i\sqrt{N}t}$), and thus 
%can be neglected in a 
%leading order approximation. This idea will be further elaborated in Sec. II.

The electron spin relaxation induced by spin-orbit interaction 
is quite slow in QDs because of the level discretization. 
\cite{Golovach_PRL_04}
Nuclear magnetic dipolar interaction gives rise to spectral 
diffusion and also leads to dephasing at a time scale of 10 
$\mu$s. \cite{Roger_PRB_03, Witzel_PRB_05, Yao_2005} 
The fluctuation of the nuclear field (the Overhauser field) 
is due to direct nuclear spin flip-flops. Generally the 
nuclear dipolar coupling is weaker than the hyperfine interaction in a 
semiconductor QD. Therefore, it would be interesting to compare the 
electron spin decoherence times caused by nuclear dipolar coupling
and the virtual processes due to hyperfine interaction in the strong
field limit. 

In this paper we present more detailed and extensive 
studies than those given in our previous paper.
\cite{Deng_PRB_06}
The remainder of the paper is organized as follows: 
in Section II, we attempt to represent the electron spin 
decoherence, especially pure dephasing, with a 
properly constructed Green's function. We set up 
the equations of motion for calculating the 
correlation function. In Section III we calculate the Green's 
functions with several different nuclear polarization for
the nuclear spin reservoir. 
The spectral functions which determine the
decay behaviors of the electron spin correlation 
functions at large time 
are found. In Section IV and V we discuss the obtained results 
and conclude the paper.  

\section{Method}

The hyperfine interaction between nuclei and electrons in the 
conduction band of a semiconductor can be 
modeled by a simple Hamiltonian, \cite{Book_NMR_A, Book_NMR_S}
\be
H = \omega_0 S^{z} + \sum_k A_k I_k^{z} S^{z} +
\sum_k \frac{A_k}{2} ( I_k^{-}S^{+} + I_k^{+}S^{-} ).
\label{ham_spin}
\ee 
%Here $\omega_0=g^*\mu_B B_0$.
Here the first term represents the Zeeman energy of the electron 
spin with $\omega_0=g^*\mu_B B_0$;
the second and last terms are the hyperfine interaction
between the electron spin and the nuclear spins. Between
the two, the second term represents the nuclear Overhauser 
field on the electron, while
the last term describes the flip-flop of the electron 
and nuclear spins. The Zeeman energy of nuclear spins are 
neglected, because the Hamiltonian in Eq. (\ref{ham_spin}) conserves
the total spin angular momentum, and a single spin nuclear
Zeeman energy is much less than that of the electron spin.
Here $A_k$ is the hyperfine coupling constant at the $k$th 
nucleus. It is proportional to the probability of the 
electron at position $\mathbf{r}$ and may be 
written as
\be\label{A_k}
A_k = A |\psi(\mathbf{r}_k)| ^2,
\ee
where $\psi(\mathbf{r})$ is the electron wavefunction
(envelope wavefunction and Bloch wavefunction at the site
of the nuclear spins). 
$A$ is the total coupling strength between the 
electron and the nuclei: $\int A(\mathbf{r})d\mathbf{r}=A$. 
In a GaAs quantum dot, $A \approx 90 ~\mu eV$. It is convenient 
to work with dimensionless units, and assume 
$A/N=1$. This means that energy is measured in units of $A/N$ 
and time is rescaled by $N/A$, with $\hbar=1$. A general 
form of $A_k$ will make the summation over $k$ quite complicated 
when looking for analytical results, especially for 
anisotropic three-dimensional quantum dots. A simple form,
which is appropriate for two-dimensional gated quantum dots 
and has been used in previous calculations, \cite{Coish_PRB_04}
\be\label{A_k_simple}
A_k = e^{-\frac{k}{N}},
\ee
is adopted in the following calculations. Here $N$ is the 
effective number of nuclear spins in the dot. This coupling 
describes a two dimensional quantum dot with electron 
wavefunction decreases as a Gaussian as the distance from the center 
of the quantum dot increases. The index $k$ labels the radial 
coordinate $r^2$. 
%Defining $A=\sum_k A_k$ and converting the 
%summation into an integral, 
$A=\int_0^{\infty} A_k dk = N$. 
In other words, $A$ and $N$ can be used interchangeably. 

The central quantity that we are going to calculate is the 
retarded Green's function of the localized spin state
\be\label{G_t}
G_{\perp}(t)=-i\theta(t) \la \Psi_0 | S^{-}(t)S^{+}(0) 
|\Psi_0 \ra .
\ee
Here $\theta(t)$ is the step function. $\Psi_0$ is the initial 
wavefunction of the system (including electron and nuclear spins),
\be \label{wave0}
|\Psi_0\ra=\left[\alpha_0 |\Downarrow \ra + 
\beta_0 |\Uparrow\ra\right] |\psi_n \ra.
\ee
We assume that the electron and nuclear spins are in a product 
state at $t=0$, and are therefore not entangled. 
Furthermore, we assume that initially 
nuclear spins are in a product state 
$|\psi_n\ra = |I_1^{z},I_2^{z},
\cdots,I_k^z,\cdots \ra$, where $I_k^z=\uparrow$ or $\downarrow$, 
i.e., nuclear spins could be either up or down at some particular 
site with a total net nuclear spin polarization $P$ that
will be defined later. Here we assume that the magnitude of 
nuclear spin is 1/2 although I=3/2 for all the isotopes of GaAs.
This assumption simplifies the algebras in the following study
though we do not anticipate this simplification to cause any 
qualitative difference in the properties of $G_{\perp}(t)$.
A mixed nuclear spin state could be
expressed in terms of a linear combination of the product 
states. Thus, the results of our calculations of product states
can be used to study more general initial nuclear states 
directly. 

An important quantity that represents the quantum coherence of 
the electron spin is the
%Equivalently, one would desire to calculate the 
quantum mechanical expectation values of transverse electron 
spin operators, 
$\la \Psi_0 | S^{-}(t)| \Psi_0 \ra$ or similarly
$\la \Psi_0 | S^{+}(t)| \Psi_0 \ra$.
The Green's function defined in Eq. (\ref{G_t}) is equivalent to 
$\la \Psi_0 | S^{-}(t)| \Psi_0 \ra$ in
the high field limit. Substituting Eq. (\ref{wave0}) into the 
expression $\la \Psi_0 | S^-(t)| \Psi_0 \ra$, we find 
\begin{eqnarray}\label{S_}
\la \Psi_0 | S^{-}(t)| \Psi_0 \ra =
\alpha_0^*\beta_0 \la \Downarrow, \psi_n | e^{iHt}S^-e^{-iHt} |
\Uparrow, 
\psi_n \ra %\\\nonumber
+ \text{higher order terms}.
\end{eqnarray}
Among higher order contributions, one or more electron spin flips 
can lead to non-zero results. 
However, as we have argued in the introduction that these 
terms are of the order
$\text{O}(1/N)$ or $\text{O}(1/N^n)$ for polarized nuclei 
in the absence of external an external magnetic field. 
The leading order contribution in Eq.(\ref{S_}) on 
the contrary does not involve any electron spin flip during the 
evolution. In fact this
term gives rise to pure dephasing (as the electron spin does not flip)
while the higher order terms involving electron spin flip
lead to decoherence 
induced by relaxation. The leading order term in Eq. (\ref{S_}) 
is exactly the Green's function defined in Eq.(\ref{G_t}) 
up to a proportional constant
 with the assumption that
$|\Psi_0 \ra = |\Downarrow,\psi_n \ra$. Another way to 
understand the pure 
dephasing term in Eq. (\ref{S_}) is to consider the phase evolution 
of the electron spin in the Sch\"{o}dinger picture,
\begin{widetext}
\be
G_{\perp}(t) &=&-i\theta(t) \langle \Downarrow; 
I_{k_1}^z,\cdots,I_{k_N}^z |
e^{iHt/\hbar}S^-e^{-itH/\hbar} S^+ |
\Downarrow;I_{k_1}^z,\cdots,I_{k_N}^z \rangle \nonumber \\
&=& -i\theta(t)
\left \{ 
\langle  \Downarrow; I_{k_1}^z,\cdots,I_{k_N}^z | e^{iHt/\hbar}
\right \} S^-
\left \{ e^{iHt/\hbar}
|  \Uparrow; I_{k_1}^z,\cdots,I_{k_N}^z \rangle 
\right \} .
\ee 
\end{widetext}
%Specifically this term describes the phase difference of the 
%two electron spin states at time $t$ due to hyperfine interaction 
%with nuclear 
%environment.
The term in the first curly brackets represents the evolution 
of the electron spin-down state in the presence of the hyperfine 
interaction, while the term in the second curly brackets 
represents the evolution of the electron spin-up state in the same
environment. If no electron spin flip occurs, any decay in the
calculated average will be due solely to dephasing between the 
electron spin-up and spin-down states. Obviously, electron spin 
flip will also cause decay of the correlation function. Therefore, 
$G_{\perp}(t)$ contains the complete decoherence information for
the electron spin in consideration.

Without loss of generality, we assume 
$|\Psi_0\ra = |\Downarrow,\psi_n \ra$ in the 
following discussions.
The equation of motion can be set up in the Heisenberg representation 
\be\label{eom_time}
i \frac{d}{dt}G_{\perp}(t) = \delta(t) -i\theta(t)
\la \Psi_0 | [S^-(t),H]; S^+(0) |\Psi_0 \ra.
\ee
The solution of this equation can be obtained
in the energy space by performing the Fourier 
transformation, after which 
we have an equation in the form of the standard equation 
of motion for the correlation 
function of two arbitrary operators
\be\label{eom}
\omega \la\la \hat{A}; \hat{B} \ra\ra_\omega
= \la \Psi_0 |\hat{A}\hat{B}|\Psi_0 \ra|_{t=0} 
+ \la\la[\hat{A},H];\hat{B}\ra\ra_\omega,
\ee
where expectation value in the first term on the right hand 
side is calculated at the initial time, and 
$\la\la \hat{A}; \hat{B} \ra\ra_\omega
\equiv\int(-i)\theta(t)\la \Psi_0 
| \hat{A}(t)\hat{B}(0)|\Psi_0\ra e^{i\omega t}dt$.

Although the hyperfine Hamiltonian [Eq. (\ref{ham_spin})] looks quite 
simple, the general solution 
has not been found. Instead one could obtain approximate 
solutions under various conditions.  
Exact solutions can be found in two simple cases: 
namely the system with 
fully polarized initial nuclear spin configuration, and the 
Hamiltonian with uniform 
coupling constants.

It is straightforward to derive the following 
equations of $G_{\perp}(\omega)$
by calculating the spin commutators repeatedly,
\begin{widetext}
\begin{eqnarray}\label{Master_Eq}
(\omega - \Omega_0)G_{\perp}(\omega) &=& 
1-\sum_k A_k \la\la n_kS^-;S^+\ra\ra_{\omega}
-\sum_k A_k\la\la I_k^-S^z;S^+\ra\ra_{\omega}, \nonumber\\
%(\omega^2 - \frac{A_k^2}{4})\la\la I_k^-;S^+\ra\ra_{\omega} &=&
%-\frac{A_k}{2}(\omega+\frac{A_k}{2})G_{\perp}(\omega)
%+ \omega A_k \la\la n_kS^-;S^+ \ra\ra_{\omega} \nonumber \\
%&+&\frac{1}{2}\sum_{k'(k)}A_kA_{k'}V_{kk'}(\omega), \nonumber \\
(\omega^2 - \frac{A_k^2}{4})\la\la I_k^-;S^+\ra\ra_{\omega} &=&
-\frac{A_k}{2}(\omega+\frac{A_k}{2})G_{\perp}(\omega)
+ \omega A_k \la\la n_kS^-;S^+ \ra\ra_{\omega}
+\frac{1}{2}\sum_{k'(k)}A_kA_{k'}V_{kk'}(\omega), \nonumber \\
(\omega^2 - \frac{A_k^2}{4})\la\la I_k^-S^z;S^+\ra\ra_{\omega} &=&
\frac{A_k^2}{4} \la\la n_kS^-;S^+\ra\ra_{\omega}
-\frac{A_k}{4}(\omega + \frac{A_k}{2})G_{\perp}(\omega) 
+ \omega\sum_{k'(k)}\frac{A_{k'}}{2}V_{kk'}(\omega),
\end{eqnarray}
\end{widetext}
with
\be
V_{kk'}(\omega)=\la\la I_k^-S^+I_{k'}^-;S^+\ra\ra_{\omega}
               -\la\la I_k^-I_{k'}^+S^-;S^+\ra\ra_{\omega},
\ee
where $n_k = \frac{1}{2} - I_k^z$ and $\Omega_0 = \omega_0 +
\frac{N}{2}$. Here $\Omega_0$ is the effective magnetic field for 
fully polarized nuclear spin reservoir. 
$\sum_{k'(k)}$ represents the summation over $k'$ for $k'\ne k$.
After rearranging the equations, we find
\begin{widetext}
\begin{eqnarray}\label{Master_Eq_1}
\left [ \omega - \Omega_0 - \Sigma_0 (\omega) \right ]G_{\perp}(\omega) &=& 
1 - \sum_k A_k\la\la n_kS^-;S^+\ra\ra_{\omega} 
- \frac{\omega}{2} \sum_{k\neq k'}\frac{A_kA_{k'}}{(\omega^2 - A_k^2/4)}
V_{kk'}(\omega) \\ \nonumber
&-& \frac{1}{4}\sum_k\frac{A_k^3}{(\omega^2-A_k^2/4)}
\la\la n_kS^-;S^+\ra\ra_{\omega}
%\nonumber \\
\end{eqnarray}
\end{widetext}
with the self-energy
\be
\Sigma_0(\omega) = \frac{1}{4} \sum_k \frac{A_k^2}{\omega - A_k/2}.
\ee
%Here $\Omega_0$ is the effective magnetic field for fully 
%polarized nuclear spin reservoir.
We will use $\Omega = \omega_0 + \sum_k A_k I_k^z$ to represent the 
effective magnetic field experienced 
by the electron in the following discussion. 
The meanings of the various correlations are obvious.
For example, $\la\la n_kS^-;S^+\ra\ra_{\omega}$
represents the Fourier transform of the time-correlation function
for flipping the electron spin to the down state while the $k$th 
nuclear spin is also in the down state ($n_k=1$). 
Both $\la\la I_k^-;S^+\ra\ra_{\omega}$ and 
$\la\la I_k^-S^z;S^+\ra\ra_{\omega}$  describe the direct
electron-nuclear-spin flip-flop. The only difference is that the 
latter one depends on the electron spin state.
$1/\Omega$ will be taken as an important expansion parameter 
in next section where 
we look for solutions of the Eqs. (\ref{Master_Eq}). 
$\Omega \sim N$ corresponds to
polarized nuclei or external magnetic field with a 
few Tesla's. $\Omega \gg 1$ will
be a good starting point for simplifying the EOMs.
$\la\la n_kS^-;S^+\ra\ra_{\omega}$ and $V_{kk'}(\omega)$ 
are higher order correlation 
functions. We should mention that the Eqs. (\ref{Master_Eq}) 
are exact. Approximations will be applied to obtain the 
general solution of arbitrary nuclear
polarization when $\Omega \sim N$. We will repeatedly use two
different approximations, large $N$ expansion and large field
expansion. These two terms have different meanings. The first 
one is just a mathematical treatment, where we simply neglect
$A_k$ when $A_k$ and $\Omega$ appear together since
$A_k$ $\sim$ 1. This approximation simplifies our EOMs 
significantly. The large field expansion,
which will only be used for solving the high frequency 
solution of the partially polarized and unpolarized nuclei, 
is a physical approximation, where we select a group of 
EOMs that describe the major contributions to the self-energies.

\section{Solutions of the Green's Function}
In this section we obtain the solutions $G_{\perp}(\omega)$ 
of Eq. (\ref{Master_Eq})
presented in Section II under various conditions. 
The real time dynamics of the Green's 
function, $G_{\perp}(t)$ can
be found using the spectral function $\rho(\omega)$,
\be \label{inv_Four}
G_{\perp}(t) = -i\theta(t) \int _{-\infty}^{\infty}
\rho(\omega) e^{-i\omega t} d\omega,
\ee
where the spectral function $\rho(\omega)$
is obtained by analytical continuation 
($\omega ~\rightarrow ~\omega+i0^+$)
using the retarded Green's function $G_{\perp}(\omega)$ 
\cite{Book_Mahan}
\be\label{spectral}
\rho (\omega) = -\frac{1}{\pi} \text{Im}G_{\perp}(\omega + i0^+).
\ee
The imaginary part of the retarded Green's function contains 
contributions from poles and branch cuts. \cite{Book_Mahan}
More specifically, the poles will 
determine the renormalized
energy or the oscillation frequencies of the electron spin 
in our case; while 
the branch cut describes the decay
of the electron spin state. Our focus in this 
section is to calculate these quantities
for various nuclear spin polarization, with or without 
external magnetic field. 

\subsection{Fully polarized nuclei reservoir}
The solution to the problem of an electron spin interacting with
a fully polarized nuclear spin reservoir is quite straightforward.
One crucial argument we will use to simplify the EOMs is 
the conservation of angular momentum at any time during
the evolution. For initially fully polarized nuclei 
(all nuclear spins in the up state) and initial wave function 
$\Psi_0=|\Downarrow,\psi_n \ra$,
this means that there is only one spin (either the electron spin 
or one of the nuclear spins) in
the down state. This restriction eliminates all the higher-order 
correlation functions present in Eq. (\ref{Master_Eq_1}). 
For example, $\la\la n_k S^-;S^+ \ra\ra_{\omega}=0$ 
because when the electron spin is flipped down, $n_k=1/2-I_k^z$ 
must be zero for fully 
polarized nuclei. $V_{kk'}(\omega)$ vanishes for the same 
reason. Therefore Eq. (\ref{Master_Eq_1}) 
can be reduced to
\be
G_{\perp}(\omega) = \frac{1}{\omega - \Omega_0 - \Sigma_0(\omega)},
\ee
where $\Omega_0 = \omega_0 + N/2$ for fully polarized nuclei.
Assuming that the hyperfine coupling constant takes the form in 
Eq. (\ref{A_k_simple}), the real and imaginary 
part of the self-energy are
\be\label{full_real}
\text{Re} \Sigma_0(\omega) = \frac{1}{4}\int_0^{\infty}
\frac{A_k^2}{\omega-\frac{A_k}{2}}dk
=-\frac{N}{2} - N\omega \text{ln}
{\left|1-\frac{1}{2\omega}\right|},
\ee
and
\be\label{full_imag}
\text{Im}\Sigma_0(\omega) = -\frac{\pi}{4}\int_0^{\infty}
A_k^2\delta\left(\omega-\frac{A_k}{2}\right)dk=
-N\pi \omega,~\text{for}~0<\omega<\frac{1}{2}.
\label{S0_RI}
\ee
In both calculations we convert the summation over $k$ 
into an integration, and use the formula
\be
\sum_k \delta\left(\omega \pm \frac{A_k}{2}\right) = \mp \frac{N}{\omega},
\ee
in finding the imaginary part. The imaginary part of the self 
energy is nonzero only
when $\frac{1}{2}>\omega >0$, this is due to the constraint
of the delta function in Eq. (\ref{S0_RI}).
The spectral function can be calculated with Eq. (\ref{spectral})
\be\label{r_full}
\rho(\omega) = \left \{
  \begin{array}{ll}
    Z_p\delta(\omega - \Omega -\text{Re}\Sigma_0) & ~ \omega\sim N \\
    -\frac{1}{N\omega}\frac{1}{\left [\frac{\omega_0}{N\omega}+\frac{1}{N}
                                +\text{ln}|1-\frac{1}{2\omega}| \right
				]^2 
+ \pi^2} 
                               & ~0 <\omega < \frac{1}{2},
   \end{array} \right.
\ee
where the renormalization factor $Z_p$ is \cite{Book_Mahan}
\be\label{renorm}
Z_p = \frac{1}{\left|1-\frac{\partial\Sigma_0(\omega)}{\partial\omega} 
\right|_{\omega_p}}.
\ee
The pole $\omega_p$ is the solution of equation 
$\omega - \Omega - \text{Re}\Sigma_0(\omega) = 0$. 
When $\omega_0 = 0$ (zero external field), 
$\omega_p \approx N/2+1/4$ and $Z_p \approx 1-1/N$.
It is clear from Eq. (\ref{r_full}) that 
there are two contributions to the spectral function 
$\rho(\omega)$, namely a pole
near $\Omega$ and a branch cut in the range of 
$[0,\frac{1}{2}]$. For a schematic, see left panel of 
Fig. (\ref{sch_spectrum}). 
In the fully polarized case, 
the weight of the pole is of order $\text{O}(1)$, 
while only a small fraction of the spectral function 
is a continuous spectrum. Since only the branch cuts 
or continuous spectrum contribute to the decoherence, 
electron spin in a fully polarized nuclear environment 
has an infinite $T_2$ time. 

In Appendix \ref{App_JW} we show that this exact solution 
can also be found with the
Jordan-Wigner representation by reducing the original spin 
Hamiltonian into the non-interacting Anderson impurity model.
\cite{Anderson_PR_61} 
The physical 
meaning from the exact solution of the Anderson model is 
that part of the time the electron spin (localized state) is 
in the spin-down state, and part of the time one 
of nuclear spins (continuous state) occupies the down state. 
The true eigenstates (resonant scattering) are the linear combinations 
of these two states.

In order to obtain real time dynamics we need to perform 
an inverse Fourier transform. This is straightforward
for the delta function part of $\rho(\omega)$, 
the contribution of the pole. However the branch cut 
integration is much more complicated, and a closed 
form cannot be found. 
\begin{figure}[t]
\begin{center}
\epsfig{file=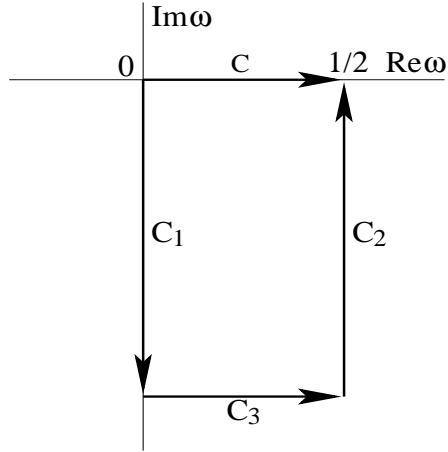, width=6cm,height=6cm}
\caption{To obtain the asymptotic behavior of $G_{\perp}(t) 
$at large time, the Fourier 
integral in Eq. (\ref{inv_Four})
is converted into a Laplace integral by deforming the 
original contour $C$ into 
$C_1+C_2+C_3$ as shown above and then allowing 
$\omega \rightarrow \infty$.}
\label{loop}
\end{center}
\end{figure}
To investigate the long time decay of the Green's function, 
the method of the steepest descent \cite{Book_Bender} is used. 
The approximate calculation of the integration is justified by 
carefully deforming the integration contour in the 
complex $\omega$ plane so that Laplace's method can be used. 
In Laplace's method the integration of an exponential 
function is approximated by the integral of this function
in the neighborhood of the global maximum of the integrand.
In order to use the method of steepest descent to calculate 
the frequency integral in Eq. (\ref{inv_Four}), we 
deform the original
integration contour $C$ into $C_1$, $C_2$, and $C_3$ 
as shown in Fig. \ref{loop}.
The integration along $C_3$ at infinity in the lower half 
plane is exponentially small in the large time limit. 
This is exactly the spirit of steepest descent, as one can 
see the term $e^{-i\omega t}$ is negligible if $\omega$ is 
analytically extended to the complex plane through 
$\omega = -is$ and $s$ is allowed to 
approach positive infinity. Substituting the spectral 
function into Eq. (\ref{inv_Four}) we obtain
\be
G_{\perp} (t) = \left[1+\text{O}\left(\frac{1}{N}\right)\right]
e^{i\frac{N}{2}t} 
- \frac{1}{N}\int^{\frac{1}{2}}_0\frac{1}{\omega}
\frac{e^{-i\omega t}d\omega}{\left[\frac{1}{N}+\text{ln}
\left|1-\frac{1}{2\omega}\right|\right]^2+\pi^2},
\ee
where we have neglected the factor $-i\theta(t)$ in 
Eq. (\ref{inv_Four}) and assumed that 
$\omega_0 = 0$. On the contour $C_1$ where $\omega = -is$, 
the integral is determined by the integration interval near $s=0^+$ 
as $t\rightarrow +\infty$ because of the term $e^{-st}$. 
By performing a Taylor expansion  around $s=0^+$ we obtain 
$\rho_{\perp}(-is) = -\frac{i}{s\text{ln}s}$. 
The asymptotic form of the integral for the inverse Fourier 
transform when $t\rightarrow \infty$ is
\be
G_{\perp}(t)=
\left[1+\text{O}\left(\frac{1}{N}\right)\right]e^{i\frac{N}{2}t} &+&
\frac{1}{N}\left[\frac{1}{\text{ln}t}+\left(\text{ln}2
-\text{ln}\gamma+i\frac{\pi}{2}\right)
\frac{1}{(\text{ln}t)^2} \right ] \nonumber \\
&+&\text{O}\left[\frac{1}{N}\frac{1}{(\text{ln}t)^3}\right],
\ee
where $\gamma = -\int_0^{\infty} \text{ln}se^{-s}ds$ is 
the Euler constant.
It can be shown that the integral along $C_3$ has a 
smaller contribution 
that is proportional to $\frac{1}{Nt\text{ln}t}$. 
The amplitude of the decay part is only of the order 
$\text{O}\left( \frac{1}{N} \right)$, 
and the leading 
order behavior at long time is $1/\text{ln}t$. We have written the 
oscillatory part as $[1+\text{O}(\frac{1}{N})]e^{i\frac{N}{2}t}$,
where we have neglected the 1/N correction of the coefficient.  
The results thus agree with those found previously. 
\cite{Khaet_03}

\subsection{One-spin-flipped nuclear reservoir}
The exact solution for the fully polarized nuclear reservoir
is instructive but of little practical relevance in current 
experimental situations, where maximum achievable nuclear 
polarization in semiconductor quantum dots is in the order of 50\%.
A pertinent question is whether there is any qualitative difference 
between partially polarized and non-polarized nuclear reservoirs
in the spectral function, spin correlation, and its decoherence
To address this question, we consider 
another simple case in which there is only one flipped nuclear 
spin in the otherwise fully polarized initial state, i.e., 
$|\Psi_0 \ra=|\Downarrow;\uparrow_1,\uparrow_2,\cdots,
\downarrow_{\bar{k}},\cdots,
\uparrow_{N} \ra$, where $\bar{k}$ is the index of the initially 
flipped nuclear spin. 

The EOMs are quite complicated even for this simple case. 
In Appendix \ref{App_EOM_One_Flipped}
we give the complete EOMs, including 
all the higher-order correlation functions.
Together with Eq. (\ref{Master_Eq_1}), these six equations form a 
closed set. The even higher-order correlation functions vanish 
because of conservation of angular momentum, as we have
discussed in the case of the fully polarized nuclear reservoir.
We point out that these equations are exact. They 
are obtained from the iterative EOMs by calculating the commutators. 
It is a formidable task to find the exact solution of 
these equations, thus we first simplify the 
equations as much as we can. It turns out to be convenient 
to discuss the low energy 
($|\omega| <\frac{1}{2}$) and 
high energy ($\omega \sim \Omega$) solutions separately. From 
the exact solution of the fully
polarized case, we already know that the spectral function 
behaves quite differently in 
these two regimes. Specifically, the low energy part corresponds 
to a continuous spectrum, while 
the high energy part is just a delta function that gives rise to 
the undamped coherent oscillation in the electron spin.
In the following we shall discuss both the low energy and high 
energy solutions of one-flipped nuclear spin reservoir with 
$1/N$ expansion. Since $N$ is very large, this 
approximation should make little difference from the exact 
solutions.

{\it Low energy solution.} To obtain the low energy solution, 
we start from 
Eq. (\ref{Master_Eq_1}).
By examining the right hand side of Eq. (\ref{Master_Eq_1}), it is 
easy to see that one needs
to find $\la\la n_k S^-;S^+\ra\ra_{\omega}$ and $V_{kk'}(\omega)$
[Recall that
$V_{kk'}(\omega)=\la\la I_k^-S^+I_{k'}^-;S^+ \ra\ra_\omega 
                -\la\la I_k^-I_{k'}^+S^-;S^+ \ra\ra_\omega $]. 
To simplify the equations, we use the fact that 
$\omega \ll \Omega$ in the low energy regime and
$A_k \ll \Omega$. As we point out in the introduction, we 
assume the existence of a large effective field so that 
$\Omega \sim N$. Therefore we can use the approximation, 
$\omega\pm\Omega\mp\frac{A_k+A_{k'}}{2} \approx \pm\Omega$.
In other words, we can neglect $\omega$ and $A_k$ whenever 
they appear in a sum with $\Omega$. This approximation helps find 
the leading order solutions.
%where corrections are of the order $\text{O}(\frac{1}{N})$. 
From Eq. (B1), we obtain
$\la\la n_kS^-;S^+ \ra\ra_\omega$,
\begin{eqnarray}\label{nkSS_one}
\la\la n_kS^-;S^+ \ra\ra_\omega = 
-\frac{1}{\Omega+\Sigma_0(\omega)}\la n_k\ra_0 
+\frac{\omega}{2\left[\Omega+\Sigma_0(\omega)\right]}
\sum_{k'(k)}\frac{A_kA_{k'}}{\omega^2-\frac{A_k^2}{4}}
V_{kk'}(\omega) %\nonumber \\
+ \text{O}\left(\frac{1}{\Omega^2}\right).
\end{eqnarray}
Here the self-energy term $\Sigma_0(\omega)$ 
is of the order $\text{O}(N)$
because it involves a summation over nuclear sites
[also see Eqs. (\ref{S0_RI}) for the fully polarized nuclear reservoir].
By studying Eq. (\ref{Master_Eq}) and the equations 
in Appendix \ref{App_EOM_One_Flipped} one can draw the conclusion 
that $V_{kk'}(\omega) \sim \la\la n_k S^-;S^+\ra\ra_\omega$.
This is because $V_{kk'}(\omega)\sim \la\la I_k^-S^-;
S^+\ra\ra_{\omega}/\Omega$, which is a result of Eq. (B4). 
By putting this relation into Eq. (\ref{Master_Eq}) one could 
immediately see that 
$V_{kk'}(\omega) \sim \la\la n_k S^-;S^+\ra\ra_\omega$.
%can be expressed in terms of $\la\la n_k S^-\;S^+\ra\ra_\omega$. 
With these results we can 
determine the order of magnitude of the summation 
$\sum_k A_k \la\la n_kS^-;S^+ \ra\ra_\omega$ 
using Eq. (\ref{nkSS_one})
\be
\sum_k A_k \la\la n_kS^-;S^+ \ra\ra_\omega \propto
\frac{1}{\Omega+\Sigma_0(\omega)}\sum_k \la n_k\ra_0 
\propto \text{O}\left( \frac{1}{\Omega}\right),
\ee
since $\sum_k \la n_k\ra_0=1$. [ Recall that there is only 
one nuclear spin in the down 
state ($n_{\bar{k}}=1$) initially, and all other nuclear spins 
point up ($n_k=0$) ]. Similarly one can
easily find that both the third and the fourth terms on the 
right hand side of 
Eq. (\ref{Master_Eq_1}) are of the order 
$\text{O}\left(\frac{1}{\Omega}\right)$. Therefore
we obtain the leading order low energy solution of 
$G_{\perp}(\omega)$ from 
Eq. (\ref{Master_Eq_1})
\be
G_{\perp}(\omega) = -\frac{1}{\Omega + \Sigma_0(\omega)} + 
\text{O}\left(\frac{1}{\Omega^2}\right),
\ee
for $\omega \sim \text{O}(1)$. This solution is exactly 
the leading order approximation of the fully
polarized nuclei in the low energy limit. Clearly there 
are only higher-order differences for the two
cases in this regime. The leading order spectral functions
also have the same behavior.

{\it High energy solution.} The approximate solution in 
the high energy limit is much more 
difficult to obtain compared to the low energy limit. 
In this limit $\omega \sim \Omega$, thus $A_k \ll \omega, \Omega$.
%$A_k \ll \omega$, and $A_k \ll \Omega$.
Generally we can drop $A_k$ in the expressions such as
$\omega \pm \frac{A_k}{2}$ and 
$\Omega \pm \frac{A_k}{2}$. However, we must be careful 
when dealing with terms like
$\omega-\Omega \pm \frac{A_k}{2}$. Since both $\omega$ and 
$\Omega$ are of order $N$, the
difference between them could be $\text{O}(1)$. 
Thus we can no longer
neglect the $A_k$ terms in these expressions.
With $\Omega \gg A_k$, we find Eq. (\ref{EOM_One_1}) can be simplified to
\begin{eqnarray}
\la\la n_kS^-;S^+\ra\ra_{\omega} = 
\frac{1}{\omega-\Omega-\Sigma_0(\omega)}
\left[\la n_k\ra_0 - \frac{1}{4\Omega}(f_k+g_k)\right],
\label{one_high_1}
\end{eqnarray}
where $f_k(\omega)=\sum_{k''}A_{k''}V_{k''k}(\omega)$ and 
$g_k(\omega)=\sum_{k''}A_{k''}V_{kk''}(\omega)$.
Substituting Eq. (\ref{one_high_1}) into Eq. (\ref{Master_Eq_1}),
we obtain
\be
G_{\perp}(\omega) = \frac{1}{\omega-\Omega-\Sigma_0(\omega)}
\left[1-\frac{1}{2\Omega}\sum_{k\neq k'}A_kA_{k'}V_{kk'}(\omega) \right].
\label{one_high_G}
\ee
Here $V_{kk'}(\omega)$ contains two terms, one of which is 
negligible compared to the 
other in the high energy limit. It can be shown that 
$\la\la I_k^-S^+I_{k'}^-;S^+\ra\ra_{\omega} 
\sim \text{O}\left( \frac{1}{\Omega}\right)
\la\la I_k^-I_{k'}^+S^-;S^+\ra\ra_{\omega}$. 
This relationship is to be expected as $\la\la I_k^-S^+I_{k'}^-;
S^+\ra\ra_{\omega}$ corresponds to the electron being flipped during
the evolution, which is a probability event 
$\left[\text{O} \left({\frac{1}{N}}\right)\right]$.
Similar simplifications will be seeked 
when we search for solutions of partially polarized and 
unpolarized nuclei.
   
In order to evaluate $V_{kk'}(\omega)$, we first simplify 
Eqs. (\ref{EOM_One_2}) and (\ref{EOM_One_5}) using the 
large $N$ expansion method 
we have outlined above.
it can be shown that the two equations can be approximated with
\be
\la\la I_k^-n_{k'};S^+\ra\ra_{\omega} = 
-\frac{A_k}{2\Omega}\la\la n_{k'}S^-;S^+\ra\ra_{\omega} + 
 \frac{A_{k'}}{2\Omega}V_{kk'}(\omega),
\label{one_high_2}
\ee
and
\be
\la\la I_k^-I_{k'}^+I_{k''}^-;S^+\ra\ra_{\omega} = 
\frac{A_k}{2\Omega}V_{k''k'}(\omega) + \frac{A_{k''}}{2\Omega}V_{kk'}(\omega).
\label{one_high_3}
\ee
Using Eqs. (\ref{EOM_One_3}), (\ref{one_high_2}), 
and (\ref{one_high_3}), we can derive an 
implicit equation for $V_{kk'}(\omega)$,
\begin{eqnarray}
\left( \tilde{\omega} + \frac{A_k+A_{k'}}{2}\right)V_{kk'}(\omega) &=&
-\frac{A_kA_{k'}}{4\Omega}\frac{\la
n_k\ra_0}{\omega-\Omega-\Sigma_0+A_k}
-\frac{A_kA_{k'}}{4\Omega}\frac{\la
n_{k'}\ra_0}{\omega-\Omega-\Sigma_0+A_{k'}}
\nonumber \\
&+&\frac{1}{4\Omega}\left[A_kg_{k'}(\omega)+A_{k'}f_k(\omega)\right],
\label{one_high_master}
\end{eqnarray}
where $\tilde{\omega}=\omega-\Omega-\frac{A_2}{4\Omega}$ and 
$A_2=\sum_kA_k^2$. Using the wavefunction in Eq. (\ref{A_k}), it is 
easy to show that $A_2=\frac{N}{2}$. Multiplying both sides of 
Eq. (\ref{one_high_master}) 
by $A_{k'}/\left[\tilde{\omega}+(A_k+A_{k'})/2\right]$ and summing 
over $k'$, we find the equation for $f_k(\omega)$ takes the form
\begin{eqnarray}
(\Omega - \alpha_k)f_k &=& 
-\frac{\alpha_kA_k\la n_k\ra_0}{\omega-\Omega-\Sigma_0+A_k}
-\frac{A_kA_{\bar{k}}^2}{4\left(\tilde{\omega}
+\frac{A_k+A_{\bar{k}}}{2}\right)
(\omega-\Omega-\Sigma_0+A_{\bar{k}})} \nonumber \\
&+& \frac{1}{4}\sum_{k''(k)}\frac{A_kA_{k''}}
{\left(\tilde{\omega}+\frac{A_k+A_{k''}}{2}\right)}g_{k''},
\label{f_k}
\end{eqnarray}
where
\be
\alpha_k(\tilde{\omega}) = \frac{1}{4}\sum_{k''}\frac{A_{k''}^2}
{\tilde{\omega}+\frac{A_k+A_{k''}}{2}}.
\ee
The equation for $g_k$ is obtained by interchanging $f_k$ and $g_k$ in 
Eq. (\ref{f_k}). 
Through the symmetry of the equations for $f_k$ and $g_k$, we find $f_k=g_k$
in the leading order large $\Omega$ expansion. 
From Eq. (\ref{f_k}) it can be seen that
$\sum_k A_k f_k \sim \text{O}(1)$ near the pole of 1/$(\omega-\Omega-\Sigma_0)$.

\begin{figure}[t]
\begin{center}
\epsfig{file=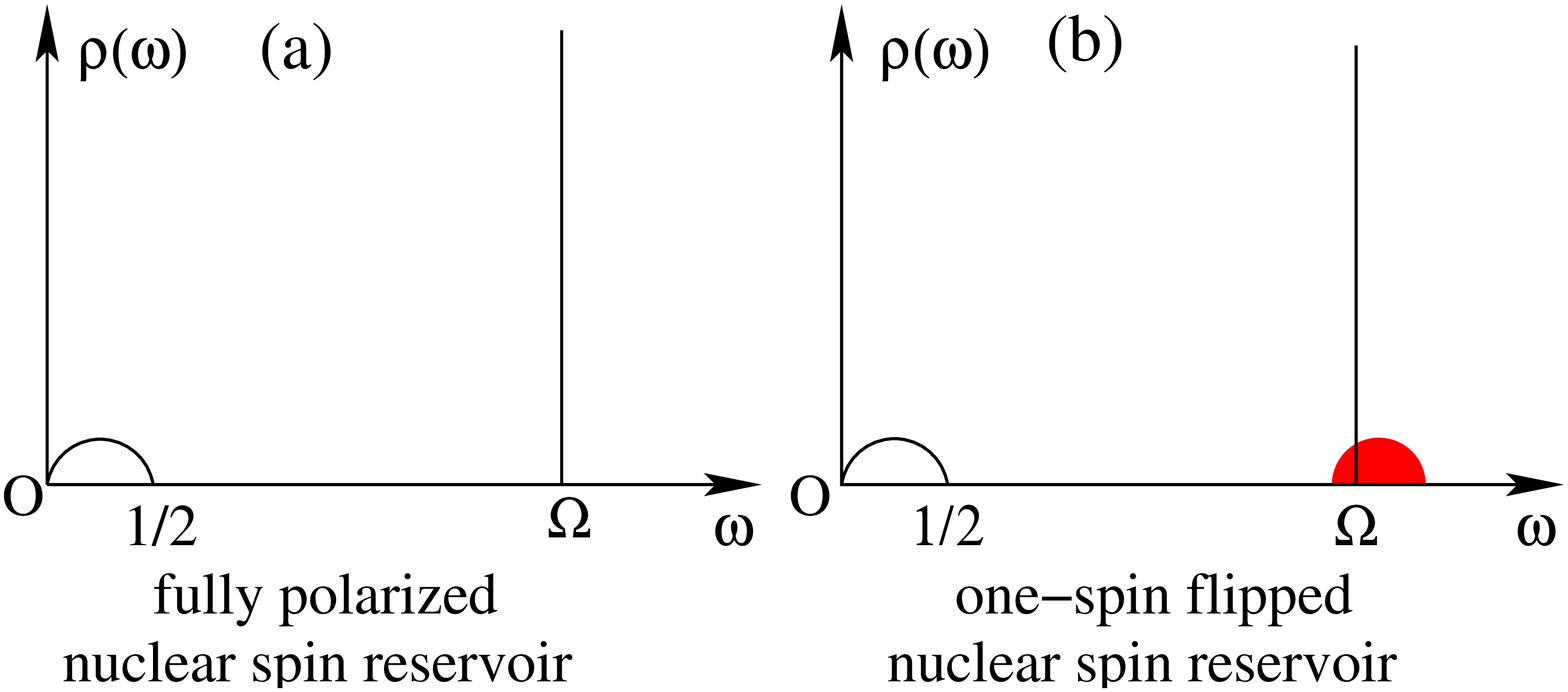, width=11cm,height=5cm}
\caption{Schematic of the spectral function for fully polarized nuclei
(lefty panel) and one-spin flipped nuclei (right panel). The shaded area
in the right figure denotes the contribution of the branch cut by 
considering virtual nuclear spin flip-flop which is not possible in
the case of fully polarized nuclei. Also the delta function represented by
the vertical line has a slightly 
reduced weight compared to that in the fully polarized case.}
\label{sch_spectrum}
\end{center}
\end{figure}

Equation (\ref{f_k}) can be
transformed to an integral equation by converting the summations 
into integrals, though this integral equation is still difficult 
to solve. For our present purpose, Eq. (\ref{one_high_G}) 
and Eq. (\ref{f_k}) are enough to further our discussions. 
First, by solving the equation 
$\omega-\Omega-\Sigma_0(\omega)=0$, we obtain the same pole as the fully 
polarized case. The interesting point here is that the pole now has a 
reduced weight, because the residue is 
$Z_p\approx 1-\frac{1}{N}-\frac{1}{2\Omega}\sum_k A_k f_k$. Since 
$\sum_kA_kf_k \sim \text{O}(1)$ when $\omega-\Omega-\Sigma_0=0$, 
the difference from the fully polarized nuclei 
is very small ($Z_p \approx 1-1/N$ in the 
exact solution). On the other hand, the function $f_k(\omega)$ 
has non-vanishing imaginary part when the frequency is in the 
neighborhood of $\Omega$. In other words, the spectral function 
has a branch cut near $\Omega$. 
This contribution to the spectral function is only of the order 
$\text{O}\left(\frac{1}{\Omega}\right)$ because of the prefactor 
$\frac{1}{2\Omega}$ in Eq. (\ref{one_high_G}). Combining these 
discussions with the picture we have established in the low 
frequency region, we can summarize the properties of the spectral
function in the right panel of Fig. \ref{sch_spectrum}.
In the left panel, we draw the schematic of the spectral 
function of the exact solution in the fully polarized case 
for comparison. It is quite 
easy to understand where the new feature comes from. In the 
fully polarized case, the only possible process is the electron-nuclei 
flip-flop (see the upper figure in Fig. \ref{flip-flop}). 
The electron flipped state is a 
high energy state, where energy $\hbar \Omega$ (neglecting the 
nuclear Zeeman energy) is needed to compensate for the 
Zeeman energy mismatch.
For a one-spin-flipped nuclei reservoir,
virtual nuclear spin flip-flop mediated by the electron is 
possible (see the lower 
figure in Fig. \ref{flip-flop}). In the initial and final states, the 
electron spin state remains unchanged, but two nuclear spins exchange 
their states. The virtual (intermediate) state is a high energy state,
and the cross section of this process is proportional to $1/\Omega$.

\subsection{Partially polarized and unpolarized nuclear
reservoir}
The solutions of both the fully polarized nuclear reservoir and 
one spin flipped nuclear reservoir, while instructive, have little 
experimental relevance. It is important to find the solution for a 
general nuclear spin configuration when studying the electron spin 
decoherence. To proceed, we first
define the effective nuclear polarization $P$ by
\be
P = \frac{N_{\uparrow}-N_{\downarrow}}{N},
\ee
where $N_{\uparrow}$ ($N_{\downarrow}$) represents the number of nuclear 
spins with up (down) spin. We further assume that both $N_{\uparrow}$ and 
$N_{\downarrow}$ are large, i.e., $N_{\uparrow}\sim N_{\downarrow}$. We can
then convert the summation over both the up and down nuclear 
spins into integrals.
We expect the solution to be similar to those we have already
found in the limit of highly polarized nuclei when 
$N_{\downarrow} \ll N_{\uparrow}$ or $N_{\downarrow} \gg N_{\uparrow}$.

\begin{figure}[t]
\begin{center}
\epsfig{file=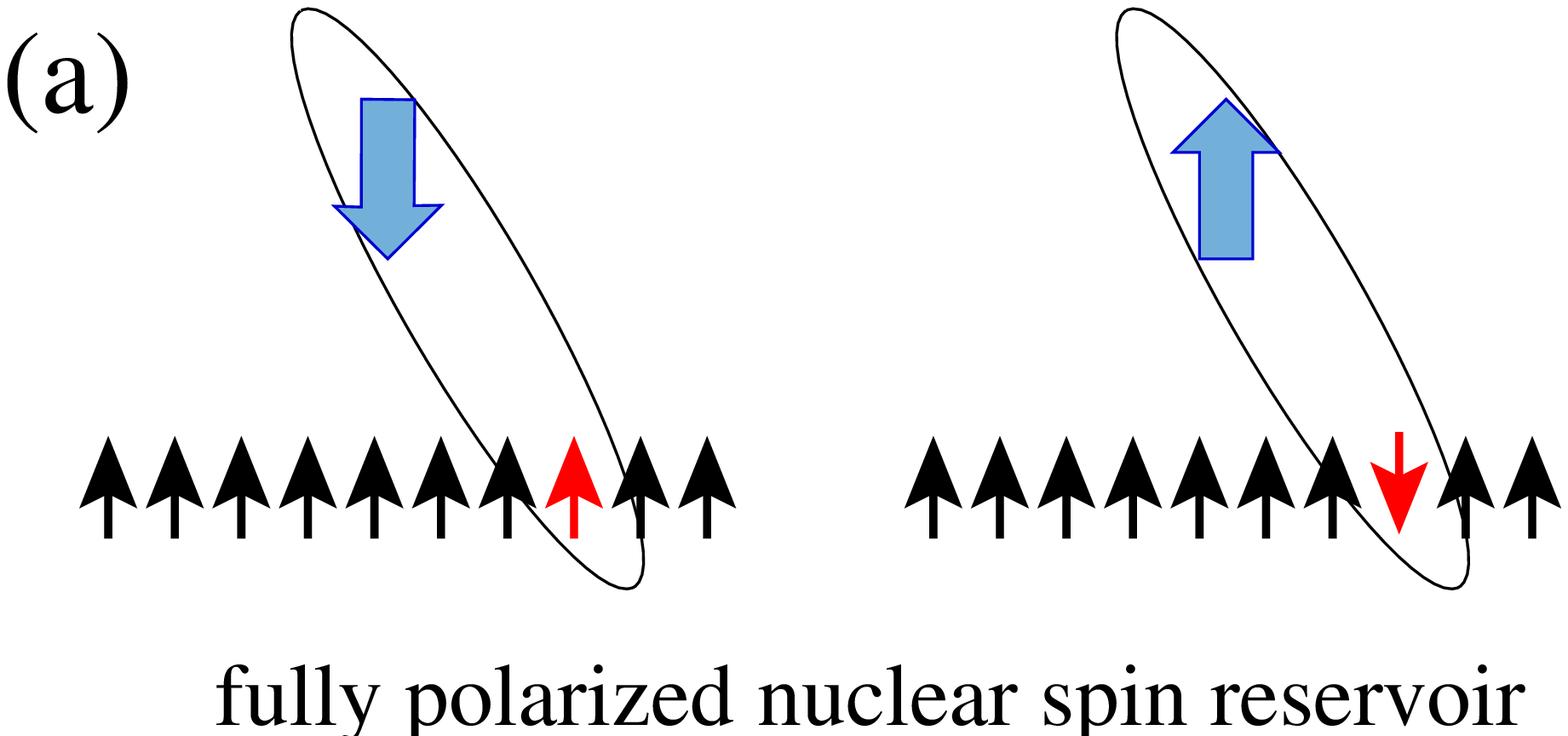, width=7.0cm,height=4cm}
\epsfig{file=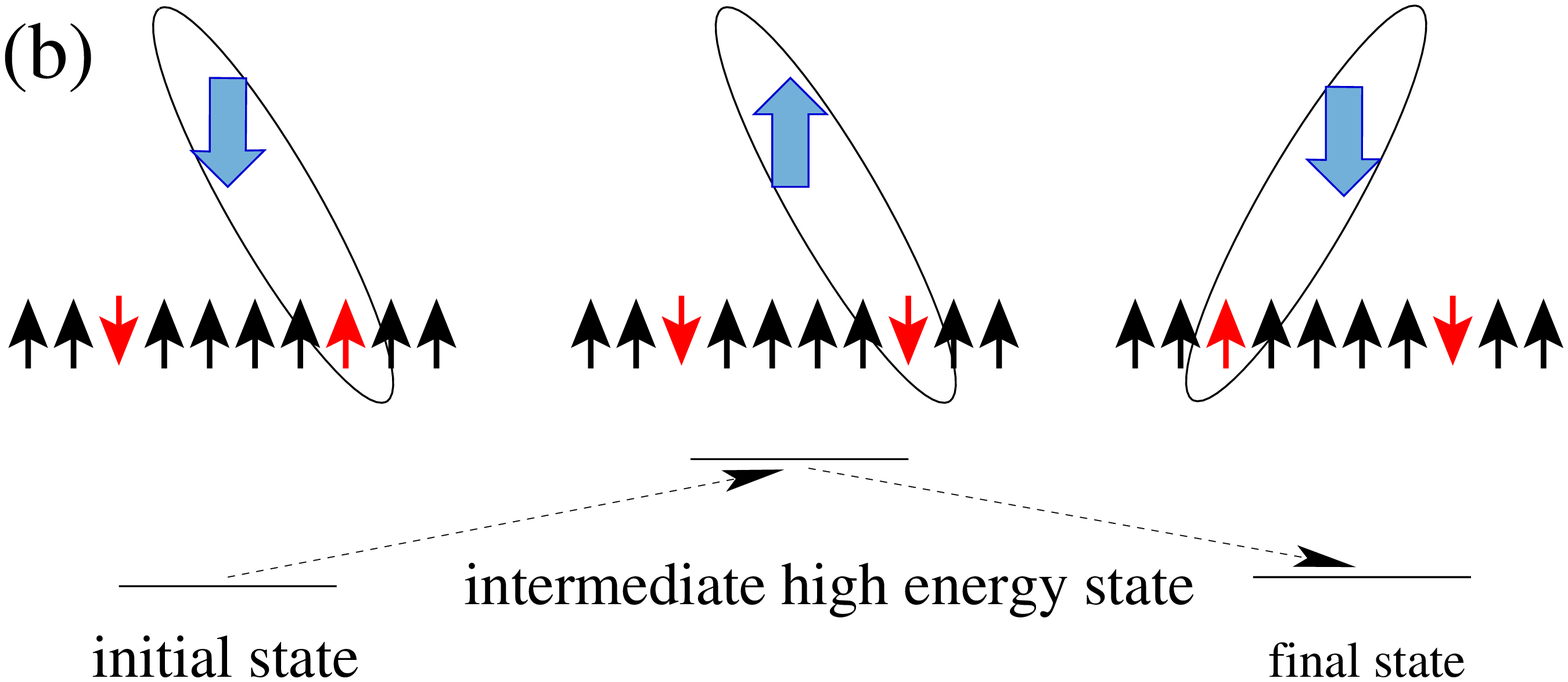, width=9.5cm,height=5cm}
\caption{Schematic of possible spin exchange processes in case of fully
polarized nuclei (upper panel) and one-spin-flipped nuclei (lower
panel). For the fully polarized nuclei, the only possible process 
is the direct electron-nuclei flip-flop, which requires a large 
assisted energy to compensate in a high magnetic field. Second-order
virtual processes are also accessible for the one-spin-flipped nuclei.
In this case, the high energy state is an intermediate state, while
the electron spin state is not changed in the initial and final states,
and two nuclear spins exchange information effectively.}
\label{flip-flop}
\end{center}
\end{figure}

We treat the nuclear field (or Overhauser field) 
$\sum_k A_k I_k^z$ with the adiabatic approximation. With a large effective
magnetic field $\Omega$, the electron spin precesses with frequency
$\frac{\Omega}{\hbar}$. On the other hand, the nuclear field fluctuates
around its average value by a small amount in a much slower time scale,
since there are many nuclear spins ($N\gg1$) interacting 
with the electron spin simultaneously. Each of the nuclear spins has only 
a $1/N$ probability to exchange spin with the electron. Under this 
approximation we can neglect the small time-dependent change of the 
nuclear field 
$\sum_k A_k I_k^z$, and factor it out from the time-dependent 
correlation function
$\la\la \sum_k A_k I_k^z(t) S^-(t);S^+\ra\ra$. Meanwhile, the nuclear spin 
raising and lowering operators are treated fully quantum 
mechanically. This approximation reduces the higher order 
correlation function $\la\la n_kS^-;S^+ \ra\ra_\omega$ 
to $G_{\perp}(\omega)$. 
%In fact this approximation  is more like a 
%mathematical treatment than a physical one. 
The term $\sum_k A_k I_k^zS^z$
in the Hamiltonian remains as an operator for calculating commutators. 
When we take the nuclear operator in the correlation functions as 
a c-number, we always make sure 
that the difference from the real quantity only gives rise to 
higher order contributions.

Similar to the highly polarized situations discussed previously,
we study the solution in the low energy and high energy limit 
separately. Recall that in the case of polarized nuclear 
reservoir with one flipped spin,
%When dealing with the solution and spectrum 
%function of one flipped nuclei, we know that 
the weight of the delta function in the spectral function 
is reduced by a small amount, and there appears a new continuous contribution 
to the spectral function near $\Omega$ because of the nuclear 
spin flip-flop mediated by the electron. Since there is only one nuclear spin
in the down state initially, scattering channels 
for nuclear flip-flops are limited (see Fig. \ref{flip-flop}). 
Therefore, decoherence effect of the 
continuous contribution is small, of order 
$\text{O}\left(\frac{1}{N}\right)$. With more nuclei in the down spin state, 
we expect that the continuous contribution will become increasingly
important, and be of the 
order unity eventually when $N_{\downarrow} \sim N_{\uparrow}$. We will 
verify this conjecture in the following discussions.

{\it Low energy solution.} 
Helped by the adiabatic approximation for the nuclear field, we can
write the first equation in Eqs. (\ref{Master_Eq}) in the following 
form,
\be
G_{\perp}(\omega) = -\frac{1}{\Omega} + 
\frac{1}{\Omega}\sum_k A_k \la\la I_k^-S^z;S^+ \ra\ra_{\omega}
+ \text{O}(\frac{1}{\Omega^2}).
\label{low_0}
\ee
Thus the leading order contribution to $G_{\perp}(\omega)$ is
$\text{O}(\frac{1}{\Omega})$. We can also write the equations for
$\la\la I_k^-;S^+\ra\ra_{\omega}$ and 
$\la\la I_k^-S^z;S^+\ra\ra_{\omega}$ as
\begin{eqnarray}
\omega\la\la I_k^-;S^+\ra\ra_{\omega} &=& 
A_k \la\la I_k^-S^z;S^+\ra\ra_{\omega} -h_kG_{\perp}(\omega), 
\label{low_1}
\\
\omega\la\la I_k^-S^z;S^+\ra\ra_{\omega} &=& -\frac{h_k}{2}G_{\perp}(\omega)
+\frac{A_k}{4}\la\la I_k^-;S^+\ra\ra_{\omega}
+ \frac{1}{2} \la\la I_k^-V;S^+\ra\ra_{\omega},
\label{low_2}
\end{eqnarray}
with $h_k = A_k\la I_k^z\ra_0$ and $V=S^+\sum_kA_k I_k^- - \sum_k A_k I_k^+S^-$.
By introducing $h_k$, we are taking a mean field average for the
Overhauser field produced by this individual nucleus.
Substituting Eqs. (\ref{low_1}) and (\ref{low_2}) into Eq. (\ref{low_0}), 
we obtain
\begin{eqnarray}
\left( \omega^2 - \frac{A_k^2}{4}\right) \la\la I_k^-S^z;S^+\ra\ra_{\omega} = 
-\frac{h_k}{2}\left(\omega+\frac{A_k}{2}\right)G_{\perp}(\omega) 
+ \frac{\omega}{2}\la\la I_k^-V;S^+\ra\ra_{\omega}.
\label{low_3}
\end{eqnarray}
Fortuitously, further calculation of $\la\la I_k^-V;S^+\ra\ra_{\omega}$ 
does not lead to any other higher order correlation functions in 
the low $\omega$ limit. This is illustrated by evaluating both 
$\la\la I_k^-S^+\tilde{I}^-;S^+\ra\ra_{\omega}$ and 
$\la\la I_k^-\tilde{I}^+S^-;S^+\ra\ra_{\omega}$ with the EOM by a 
$1/N$ expansion.
Here we have defined $\tilde{I}^{\pm}_n=\sum_k A_k^n I_k^{\pm}$ as 
collective nuclear spin operators. The index is dropped when $n=1$.
Neglecting all higher order terms, we find
\begin{eqnarray}
&&\Omega\la\la I_k^-S^+\tilde{I}^-;S^+\ra\ra_{\omega} =
-h_k \la\la I_k^-\left(\frac{1}{2}-S^z\right);S^+\ra\ra_{\omega} \nonumber \\
&-&\la\tilde{I}^z_2\ra_0 
\la\la I_k^-\left(\frac{1}{2}-S^z\right);S^+\ra\ra_{\omega} 
+ \la\la I_k^-\tilde{I}^+\tilde{I}^-;S^+\ra\ra_{\omega}, \nonumber \\
&&\Omega\la\la I_k^-\tilde{I}^+S^-;S^+\ra\ra_{\omega} =
-\la I_k^-\tilde{I}^+\left(\frac{1}{2}-S^z\right)\ra_0 \nonumber \\
&-&\la\tilde{I}^z_2\ra_0 
\la\la I_k^-\left(\frac{1}{2}-S^z\right);S^+\ra\ra_{\omega} 
+ \la\la I_k^-\tilde{I}^+\tilde{I}^-;S^+\ra\ra_{\omega}, 
\label{low_V}
\end{eqnarray}
where we have used the operator equality $S^+S^- = \frac{1}{2} - S^z$ 
and the fact
that $\la\la I_k^-S^+\tilde{I}^-;S^+ \ra\ra_{\omega} \sim 
\text{O}\left[\la\la I_k^-S^+\tilde{I}_2^-;S^+\ra\ra_{\omega}\right]$. 
Combining Eq. (\ref{low_1}) with (\ref{low_V}), 
we find a single equation for $\la\la I_k^-S^z;S^+\ra\ra_{\omega}$
and $G_{\perp}(\omega)$,
\begin{eqnarray}
\left(\omega^2 - \frac{A_k^2}{4}\right)
\la\la I_k^-S^z;S^+\ra\ra_{\omega} &=&
\frac{\omega A_k}{4\Omega} - \frac{PA_k^2}{8}G_{\perp}(\omega) 
+ \frac{P^2NA_k}{32\Omega}G_{\perp}(\omega) \nonumber \\
&+& \frac{PA_k}{8\Omega}\sum_{k'} A_{k'}^2 \la\la I_{k'}^-S^z;S^+\ra\ra_{\omega}.
\label{low_Master}
\end{eqnarray}
In this equation, we have used the mean field average and,
replaced the nuclear spin expectation value of
the $z$ component $\la I_k^z\ra_0$ by the average nuclear polarization 
$\frac{P}{2}$.
The summation $\sum_k A_k^2 \la\la I_k^-S^z;S^+\ra\ra_{\omega}$ 
in Eq. (\ref{low_Master}) can be found by multiplying both sides of 
the equation by $\frac{A_k^2}{\omega^2-A_k^2/4}$ and then 
performing the summation over $k$. We then obtain
\be
\sum_k A_k^2 \la\la I_k^-S^z;S^+\ra\ra_{\omega} = 
\frac{8\omega\sigma_3 + (P^2N\sigma_3-4P\Omega\sigma_4)G_{\perp}}
{32\Omega + 4P\sigma_3},
\label{low_sum}
\ee
with
\be
\sigma_n(\omega) = \sum_k \frac{A_k^n}{\omega^2 - \frac{A_k^2}{4}}.
\label{self_low}
\ee
$\sigma_n \sim N$ since it is a summation over all the nuclear spins.
Plugging Eq. (\ref{low_sum}) back into
Eq. (\ref{low_Master}), we obtain an expression for 
$\sum_k A_k \la\la I_k^-S^z;S^+\ra\ra_{\omega}$, which can then be used to find 
$G_{\perp}(\omega)$ using Eq. (\ref{low_0}). After all these evaluations, 
we obtain the low energy solution for $G_{\perp}(\omega)$:
\be
G_{\perp}(\omega) = -8
\frac{8\Omega-2\omega\sigma_2+P\sigma_3}
{(8\Omega+P\sigma_3)^2-P^2\sigma_2(2N+\sigma_4)}.
\ee
The final solution is quite complicated. One important result is that 
$G_{\perp}(\omega) \sim \text{O}(\frac{1}{N})$, just like in the highly 
polarized cases. In this 
solution the higher-order correlation function has no effect since the 
term $\la\la I_k^-\tilde{I}^+\tilde{I}^-;S^+\ra\ra_{\omega}$, which 
involves nuclear spin flip-flop, does not appear in Eq. (\ref{low_Master}).
The summation $\sigma_n(\omega)$ can be calculated with the same technique as 
we have used for calculating $\Sigma_0(\omega)$ with analytical continuation
$(\omega \rightarrow \omega + i0^+)$ 
and the conversion of summation to integral
$(\sum_k \rightarrow \int_0^{\infty} dk)$. The real and imaginary parts of
$\sigma_n$ are evaluated and given in Appendix \ref{App_self_low}. 
The real parts of 
$\sigma_n$ have two branch cuts, one from 0 to $\frac{1}{2}$; 
the other one from 
$-\frac{1}{2}$ to 0. The imaginary part of $G_{\perp}(\omega)$ is non-zero only
in these regions.

\begin{figure}[t]
\begin{center}
\epsfig{file=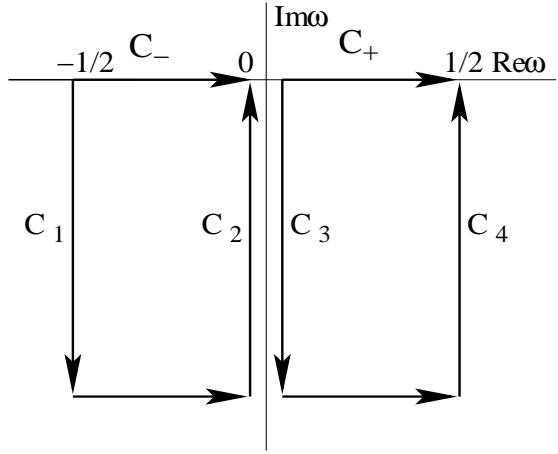, width=7.5cm,height=6cm}
\caption{To obtain the asymptotic behavior of $G_{\perp}(t)$ for
partially polarized nuclei at long time limit, the Fourier integral in 
Eq. (\ref{inv_Four}) is converted into a Laplace integral by deforming 
the original contour $C_+$ and $C_-$ into $C_1$, $C_2$, $C_3$, and
$C_4$ as shown above and then allowing $\omega~\rightarrow~\infty$.}
\label{loop_1}
\end{center}
\end{figure}

We first consider the case of zero external magnetic field, when 
$\Omega=\frac{PN}{2}$. To study the long time behavior we again deform
the original integration contours $C_+$ and $C_-$ into $C_1$, $C_2$,
$C_3$, and $C_4$ in Fig. \ref{loop_1}. The integrals at minus infinity 
in the lower half plane are neglected. Using the results of 
$\sigma_n(\omega)$ given in the Appendix \ref{App_self_low}, 
we find the spectral function behaves as
\be
\rho(0^{\mp}-is) = \frac{i(P\pm 1)}{2NP}\frac{1}{s\text{ln}s},
\ee
on contour $C_2$ ($\omega=0^- - is$) and $C_3$ ($\omega=0^+-is$) as 
$s$ approaches $0^+$. 
Calculating the Laplace transform, we then obtain the long time asymptotic
form of time dynamics of the Green's function in the low energy limit,
\be
G_{\perp}(t) \propto \frac{1}{PN}\frac{1}{\text{ln}t},~t\rightarrow \infty.
\ee
The long time asymptotic form $1/\text{ln}t$ is the same as that of the
fully polarized case, and we do have the right numerical limit as 
$P\rightarrow 1$.
Consider now the situation of the strong Zeeman field limit where 
$|\omega_0|\gg N$. The 
Green's function is proportional to $\frac{1}{\Omega}$ when 
$\omega \ll N$. The 
asymptotic behavior of $G_{\perp}(t)$ will be $\frac{1}{t}$, and the 
decaying amplitude is 
proportional to $\frac{1}{\Omega}$. Again this result is the same 
as the fully polarized nuclei case. \cite{Khaet_03, Coish_PRB_04}

The properties of the low energy solution to the spin correlation 
function studied in this 
subsection are quite similar to those that have been investigated in 
previous studies. \cite{Khaet_03,Coish_PRB_04} The basic 
result is that the decay amplitude is of the order 
$\text{O}\left( \frac{1}{N}\right)$ in the perturbative limit when the 
effective magnetic field is comparable to the number of nuclear spins
in the QD. However, only the direct electron-nuclei flip-flop has been
considered in these studies. Therefore it is quite reasonable that
the contribution is small since the Zeeman energy mismatch is large. 
%In the 
%following discussion of the solution in the high frequency region, we
%treat the indirect nuclear spin exchange systematically.

{\it High energy solution.} 
In the following discussion we search for the solution to the spin 
correlation function in the high frequency region. 
%we treat the indirect nuclear spin exchange systematically.
This solution will be of the most experimental relevance. 
In the case of a highly polarized nuclear reservoir
the major contribution to the spectral 
function in the large $\omega$ limit is a delta function, 
that does not lead to 
any decoherence in the electron spin. When the effect of the electron 
mediated nuclear spin exchange is 
included, decoherence with a small amplitude arises. The question 
is whether this effect will become large enough so that there is a 
measurable contribution to electron spin dephasing when initially 
the two nuclear spin species (up and down)
have approximately equal population. We address this question in the 
following discussion of the high energy solution for partially 
polarized and unpolarized nuclei.
%The result presented below will be the major contribution of this paper.

%We again start from Eq. (\ref{Master_Eq_1}) 
By using the adiabatic approximation and 
considering that $\omega = \Omega + \text{O}(1)$ and $\omega \gg A_k$, 
Eq. (\ref{Master_Eq_1}) can be straightforwardly simplified to 
\begin{eqnarray}
\left(\omega - \Omega - \frac{A_2}{4\Omega}
\right)G_{\perp}(\omega) = 1 
- \frac{1}{2\Omega}\sum_{k\neq k'}A_kA_{k'}
\la\la I_k^-I_{k'}^+S^-;S^+\ra\ra_{\omega},
\label{high_0}
\end{eqnarray}
where $A_2 = \sum_k A_k^2$.
Notice that 
$\la\la I_k^-S^+I^-_{k'};S^+\ra\ra_{\omega} \ll 
 \la\la I_k^-I^+_{k'}S^-;S^+\ra\ra_{\omega}$, because 
$\la\la I_k^-I^+_{k'}S^-;S^+\ra\ra_{\omega}$ is proportional to 
$\frac{1}{\omega-\Omega+(A_k-A_{k'})/2}$, while 
$\la\la I_k^-S^+I^-_{k'};S^+\ra\ra_{\omega}$ is proportional to 
$\frac{1}{\omega+\Omega}$. Mathematically, the difference originates 
from the commutators of $S^{\pm}$ with
$S^z$, i.e., $[S^{\pm},S^z]=\mp S^{\pm}$. Since $\omega = \Omega+\text{O}(1)$, 
it is clear that $\frac{1}{\omega-\Omega+(A_k-A_{k'})/2} \sim \text{O}(1)$ 
while $\frac{1}{\omega+\Omega} \sim \text{O}(\frac{1}{2\Omega})$. 
Physically this difference is also easy to understand. 
In $\la\la I_k^-S^+I^-_{k'};S^+\ra\ra_{\omega}$ an electron 
spin needs to be flipped down, while in 
$\la\la I_k^-I^+_{k'}S^-;S^+\ra\ra_{\omega}$, the
electron spin is flipped up and then flipped down, and two 
nuclear spins at $k$ and $k'$
flip-flop with each other. Therefore the process in
$\la\la I_k^-I^+_{k'}S^-;S^+\ra\ra_{\omega}$
does not require a real electron spin flip and is a nuclear
flip-flop process mediated by the electron. From Eq. (\ref{high_0}) 
it is clear that this is the only higher-order correlation function 
that contribute to $G_{\perp}(\omega)$.

The calculation of the one-pair correlation function 
$\la\la I^-_k I_{k'}^+S^-;S^+\ra\ra_{\omega}$ is non-trivial, since 
its computation involves 
two-pair correlation function 
$\la\la I_{k_1}^-I_{k_2}^+I_{k_3}^-I_{k_4}^+S^-;S^+\ra\ra_{\omega}$, 
which in turn depends on even higher order correlation functions. 
We thus need a cut-off scheme to close the EOM. This is not a problem for
the low energy solution where the terms of the one-pair correlation 
functions cancel each other so that the transverse electron spin 
Green's function only depends on
the electron-nuclei spin-flip correlation functions 
$\la\la I_k^-;S^+\ra\ra_{\omega}$ and $\la\la I_k^-S^z;S^+\ra\ra_{\omega}$. 
In the high energy regime, it turns out that the self-energy function 
can be expanded in powers of $\frac{N}{4\Omega}$.
Therefore, as long as the effective field $\Omega$ is sufficiently large 
compared to $N$ (meaning that the applied field is larger than the 
Overhauser field from a fully polarized nuclear spin reservoir), the 
approximate solution can be obtained with the first few terms in the 
expansion.

We start with the exact EOM for
$\la\la I_k^-I_{k'}^+S^-;S^+\ra\ra_{\omega}$,
\begin{eqnarray}
&&\left( \omega - \Omega + \frac{A_k-A_k'}{2}\right)
\la\la I_k^-I_{k'}^+S^-;S^+\ra\ra_{\omega} = 
h_{k'}\la\la I_k^-\left( \frac{1}{2}+S^z\right)
;S^+\ra\ra_{\omega} \nonumber \\
&-&A_{k'}\left(\frac{1}{2}+I_{k'}^z\right)
\la\la I_k^-S^z;S^+\ra\ra_{\omega}
-\la\la I_k^-I_{k'}^+\sum_{k''(k,k')}A_{k''}I_{k''}^-S^z;S^+\ra\ra_{\omega}.
%-\sum_{k''(k,k')}A_{k''}
%\la\la I_k^-I_{k'}^+I_{k''}^-S^z;S^+\ra\ra_{\omega}.
\label{high_1}
\end{eqnarray}
The leading order (in terms of $\frac{1}{\Omega}$) EOM for 
$\la\la I_k^-I_{k'}^+\sum_{k''(k,k')}A_{k''}I_{k''}^-S^z;S^+\ra\ra_{\omega}$
takes the form
\begin{eqnarray}
&&\omega \la\la I_k^-I_{k'}^+
\sum_{k''(k,k')}A_{k''}I_{k''}^-S^z;S^+\ra\ra_{\omega}
= -\frac{A_2}{4}\la\la I_k^-I_{k'}^+S^-;
S^+\ra\ra_{\omega} \nonumber \\
&-&\frac{A_k}{4}\la\la I_{k'}^+\sum_{k''(k,k')}
I_{k''}^-S^-;S^+\ra\ra_{\omega},
\label{high_2}
\end{eqnarray}
where two-pair correlation functions have been neglected. 
In leading order of 1/$\Omega$, the EOMs of the two lowest-order
correlation functions of electron-nuclei flip-flopping 
in Eq. (\ref{Master_Eq}) are simplified to
\be
\la\la I_k^-;S^+\ra\ra_{\omega} = -\frac{P}{2\Omega}A_k G_{\perp}(\omega),
\label{high_3}
\ee
and
\be
\la\la I_k^-S^z;S^+\ra\ra_{\omega} = -\frac{A_k}{4\Omega}G_{\perp}
-\frac{1}{2\Omega}\la\la I_k^-\sum_{k'(k)} I_{k'}^+S^-;S^+\ra\ra_{\omega},
\label{high_4}
\ee
Combining Eq. (\ref{high_1}) with Eq. (\ref{high_4}) we find the 
following equation for $\la\la I_k^-I_{k'}^+S^-;S^+\ra\ra_{\omega}$,
\begin{eqnarray}
&&\left( \tilde{\omega} + \frac{A_k-A_{k'}}{2}\right)
\la\la I_k^-I_{k'}^+S^-;S^+\ra\ra_{\omega} = 
\frac{1-P^2}{8\Omega}A_kA_{k'}G_{\perp}(\omega) \nonumber \\
&+& \frac{A_{k'}}{4\Omega} 
\la\la I_{k}^-\sum_{k''(k,k')}A_{k''}I_{k''}^+S^-;S^+\ra\ra_{\omega}
%\nonumber \\
+ \frac{A_k}{4\Omega}
\la\la I_{k'}^+\sum_{k''(k,k')}A_{k''}I_{k''}^-S^-;S^+\ra\ra_{\omega}.
\label{high_master}
\end{eqnarray}
with $\tilde{\omega} = \omega - \Omega - \frac{A_2}{4\Omega}$, which 
represents the deviation of the frequency from $\Omega$ in the high 
energy limit. Note that $A_2 /\Omega \sim \text{O}(1)$. Equation 
(\ref{high_master}) is hard to solve 
because of the summation on the right hand side of the equation. 
We thus look for an approximate solution in the leading order of 
$\frac{N}{\Omega}$ in the large
$\Omega$ limit ($\Omega \gg N$). It is clear by examining the equation that the
last two terms on the right hand side of the equation makes a higher order 
contribution, and can be neglected in a large field expansion.
Dividing both sides of Eq. (\ref{high_master}) by
$\tilde{\omega}-\frac{A_k-A_{k'}}{2}$ and performing the summation over $k'$, 
we obtain
\begin{eqnarray}
\la\la I_k^-\tilde{I}^+S^-;S^+ \ra\ra_{\omega} =
\left[ \frac{1-P^2}{8\Omega}A_k\sum_{k'}
\frac{A_{k'}^2}{\tilde{\omega}+\frac{A_k-A_{k'}}{2}} \right. 
\left. + \text{O}\left(\frac{N^2}
{16\Omega^2}\right)\right]G_{\perp}(\omega).
\label{high_5}
\end{eqnarray}
One can solve for $\la\la \tilde{I}^-I_{k'}^+S^-;S^+\ra\ra_{\omega}$ 
following a similar approach,
\begin{eqnarray}
\la\la \tilde{I}^-I_{k'}^+S^-;S^+ \ra\ra_{\omega} =
\left[ \frac{1-P^2}{8\Omega}A_{k'}\sum_{k}
\frac{A_{k}^2}{\tilde{\omega}+\frac{A_k-A_{k'}}{2}} \right.
\left. + \text{O}\left(\frac{N^2}
{16\Omega^2}\right)\right]G_{\perp}(\omega).
\label{high_6}
\end{eqnarray}
We then substitute Eqs. (\ref{high_5}) and (\ref{high_6}) into 
Eq. (\ref{high_master}) 
to obtain the solution for $\la\la I_k^-I_{k'}^+S^-;S^+\ra\ra_{\omega}$. 
Finally we use Eq. (\ref{high_0}) to find the Green's function in the 
high energy limit. The result is
\be
G_{\perp}(\omega) = \frac{1}{\tilde{\omega} 
- \frac{1-P^2}{16\Omega^2}\Sigma_1(\tilde{\omega})
- \frac{1-P^2}{64\Omega^3}\left[ \Sigma_2(\tilde{\omega}) 
                               + \Sigma_3(\tilde{\omega})\right]
\label{high_G}
}
\ee
where
\be
\Sigma_1(\tilde{\omega}) = \sum_{k,k'}
\frac{A_k^2A_{k'}^2}{\tilde{\omega}+\frac{A_k-A_{k'}}{2}},
\label{Self_1}
\ee
\be
\Sigma_2(\tilde{\omega}) = \sum_{k,k',k''}
\frac{A_k^2A_{k'}^2A_{k''}^2}
{(\tilde{\omega}+\frac{A_k-A_{k'}}{2})(\tilde{\omega}+\frac{A_k-A_{k''}}{2})},
\label{Self_2}
\ee
and
\be
\Sigma_3(\tilde{\omega}) = \sum_{k,k',k''}
\frac{A_k^2A_{k'}^2A_{k''}^2}
{(\tilde{\omega}+\frac{A_k-A_{k'}}{2})(\tilde{\omega}+\frac{A_{k''}-A_{k'}}{2})}.
\label{Self_3}
\ee
Because $\Sigma_1({\tilde{\omega}})$ has two summations over nuclear spins, 
$\Sigma_1({\tilde{\omega}}) \sim N^2$. For the same reason, 
$\Sigma_2(\tilde{\omega})$ and $\Sigma_3(\tilde{\omega})$ are proportional to
$N^3$. In the above derivation, we have neglected the difference between
$\sum_{k''(k,k')}$ and $\sum_{k''}$, which only gives an order 
unity contribution
to the three self-energies. The final expression of $G_{\perp}(\omega)$ 
we have found in Eq. (\ref{high_G}) is of order $\text{O}(1)$, 
in contrast to the solution in the low-energy limit.

\begin{figure}
\begin{center}
\epsfig{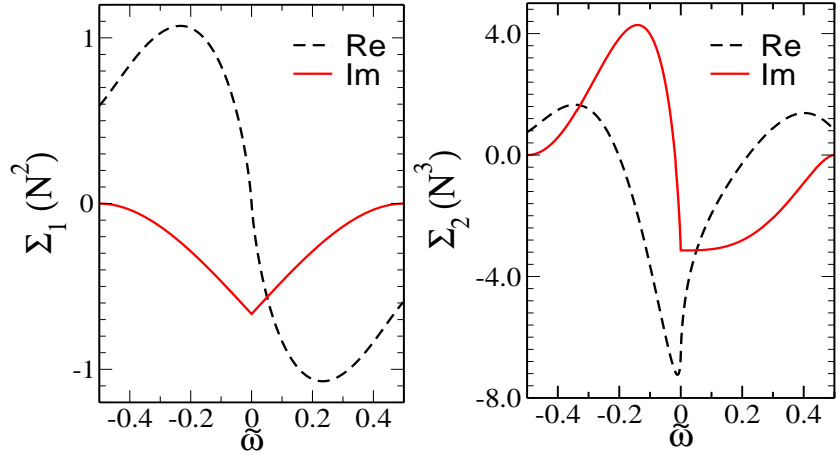}
\caption{The self-energies $\Sigma_1(\tilde{\omega})$ (left panel) 
and $\Sigma_2(\tilde{\omega})$ (right panel) as functions of
$\tilde{\omega}$ = $\omega$ - $\Omega$ - $A_2/\Omega$. In these
numerical calculations, we have assumed that the hyperfine coupling
constant takes the form as in Eq. (\ref{A_k}). Both the real part
(dashed line) and imaginary part (solid line) of the self-energies
are plotted.}
\label{Self}
\end{center}
\end{figure}

The evaluation of self-energies $\Sigma_1(\tilde{\omega})$, 
$\Sigma_2(\tilde{\omega})$, and 
$\Sigma_3(\tilde{\omega})$ is quite complicated. 
In Appendix \ref{App_self_high} we first show that
$\Sigma_2(\tilde{\omega})=\Sigma_3(\tilde{\omega})$.
After converting the summations 
into integrals, we then perform two-dimensional and three-dimensional 
integrations for $\Sigma_1(\tilde{\omega})$ and $\Sigma_2(\tilde{\omega})$.
%This is done in Appendix \ref{App_self_high}. 
In Fig. \ref{Self}, we plot the real and 
imaginary parts for both of the
self-energy terms as a function of the shifted frequency $\tilde{\omega}$. We 
can see that there is a cusp at $\tilde{\omega}=0$ for the imaginary 
parts of both 
$\Sigma_1(\tilde{\omega})$ and $\Sigma_2(\tilde{\omega})$. Let us first 
take a closer look at $\Sigma_1(\tilde{\omega})$. Notice that 
$\text{Im}\Sigma_1(\tilde{\omega})$
is negative for the whole region of $\tilde{\omega}$. This ensures 
that the spectral function $\rho(\tilde{\omega})$ is always 
positive. Like $\text{Im}\sigma_n(\omega)$ [see Eq. (\ref{self_low}) 
and Appendix C], which is nonzero when 
$-\frac{1}{2}<\omega<\frac{1}{2}$,
$\text{Im}\Sigma_1(\tilde{\omega})$ does not vanish when 
$-\frac{1}{2}<\tilde{\omega}<\frac{1}{2}$. 
%The real part of $\Sigma_1(\tilde{\omega})$,
%which renormalizes the electron spin coherent oscillation frequency, 
%has peaks at $\tilde{\omega} \approx \pm 0.25$.
The behavior of $\Sigma_2(\tilde{\omega})$ is very different from 
$\Sigma_1(\tilde{\omega})$. Neither the real 
nor the imaginary part of $\Sigma_2(\tilde{\omega})$ is symmetrical 
about $\tilde{\omega}=0$. On the other hand, 
Im$\Sigma_2(\tilde{\omega})$ changes sign near $\tilde{\omega}=0$. 
This leads to no contradiction because $\Sigma_2(\tilde{\omega})$ 
is a higher order 
correction term to $\Sigma_1(\tilde{\omega})$, and
$\Sigma_2(\tilde{\omega})$ by itself has no physical meaning. 
%One should not be 
%confused by he absolute values of $\Sigma_1(\tilde{\omega})$ and 
%$\Sigma_2(\tilde{\omega})$ 
%that we have shown in 
Fig. \ref{Self} shows that the absolute values of 
$\Sigma_2(\tilde{\omega})$ are generally larger than those of 
$\Sigma_1(\tilde{\omega})$, these self-energy terms 
are multiplied by $\frac{N^2}{16\Omega^2}$ and $\frac{N^3}{64\Omega^3}$, 
respectively, in Eq. (\ref{high_G}), so that $\Sigma_1(\tilde{\omega})$
is still the leading order contribution.

\begin{figure}
\begin{center}
\epsfig{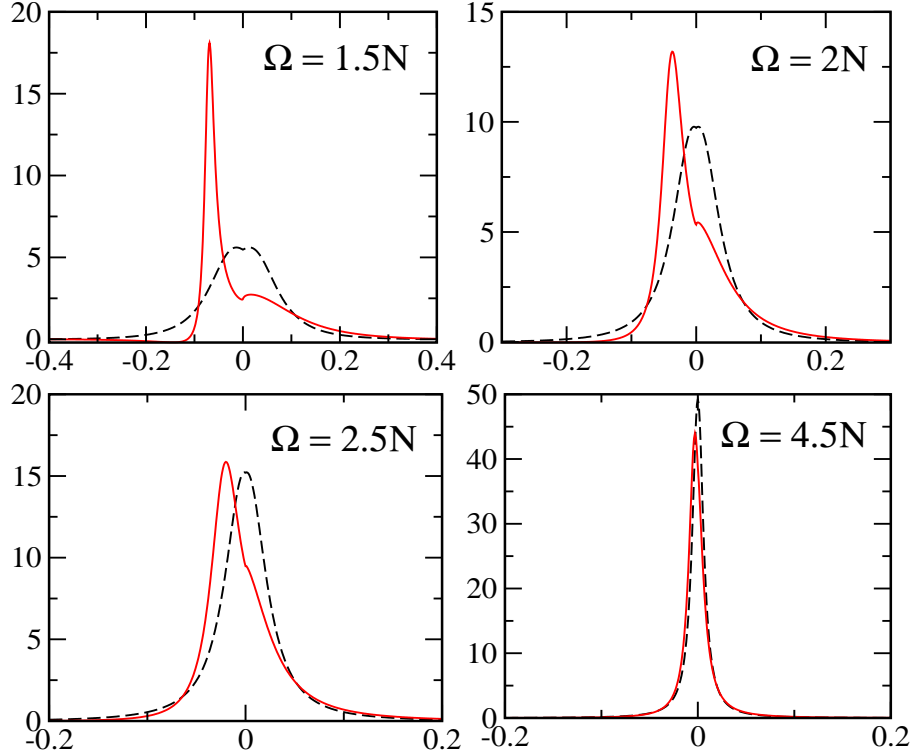}
\caption{The spectral function $\rho(\tilde{\omega})$ in the high energy limit
as a function of the shifted frequency $\tilde{\omega}$. We have 
plotted the spectral function for $\Omega$ = 1.5$N$, $\Omega$ = 2$N$,
$\Omega$ = 2.5$N$, and $\Omega$ = 4.5$N$. In each of these cases, we
have calculated $\rho(\tilde{\omega})$ with two approximations. In 
the first one, we have evaluated $\rho(\tilde{\omega})$ with only
$\Sigma_1(\tilde{\omega})$ (dashed line); and in the other one, we 
have used both $\Sigma_1(\tilde{\omega})$ and $\Sigma_2(\tilde{\omega})$ 
(solid line). The lower figures clearly show the success of our large
field expansion method. These plots also illustrated the idea that the
original delta function in the spectral function is broadened to a 
continuous spectral function after the virtual nuclear spin flip-flop
is included in the calculations.}
\label{Spectrum}
\end{center}
\end{figure}

So far the two-pair correlation function has been 
neglected when deriving the EOMs. If we 
keep the two-pair correlation function and all the other 
corresponding functions of the same order, but neglect the 
three-pair correlation functions, we would find two more higher-order 
self-energy terms which are proportional to 
$\frac{N^4}{(4\Omega)^4}$ and $\frac{N^5}{(4\Omega)^5}$. In principle, we 
can perform the large field expansion to higher and higher orders, 
and eventually, we should arrive at a finite geometrical series of 
$\frac{N}{4\Omega}$. We would get a finite instead of an infinite 
series because there is a finite number ($N$) of nuclear spins in 
the quantum dot, and there can be at most $N_{\downarrow}$ ($N_{\uparrow}$) 
pairs of flip-flopping nuclear spins if $N_{\downarrow}<N_{\uparrow}$ 
($N_{\uparrow}<N_{\downarrow}$). 
This gives rise to a natural cut-off of self-energy terms in our large 
field expansion. However, even extension 
of the calculation to the order of $\frac{N^4}{(4\Omega)^4}$ and 
$\frac{N^5}{(4\Omega)^5}$ is already very complicated. Nevertheless, 
the approximation with the first two self-energy terms should be accurate 
enough as long as the total effective magnetic field $\Omega$ is large 
compared with $N$ (representing the Overhauser field produced by a
polarized nuclei spin reservoir). 

The validity of our large field expansion method  is 
illustrated in Fig. \ref{Spectrum}, where we have computed and 
plotted the spectral function $\rho(\tilde{\omega})$ for various 
$\Omega$. To compare, we have calculated $\rho(\tilde{\omega})$ 
using only the self-energy $\Sigma_1(\tilde{\omega})$ and both 
$\Sigma_1(\tilde{\omega})$ and $\Sigma_2(\tilde{\omega})$. In the 
two upper panel figures, where $\Omega$ is not very large, the 
contribution of $\Sigma_2(\tilde{\omega})$ is comparable to 
$\Sigma_1(\tilde{\omega})$. In this case, one needs to calculate 
more higher order terms of self-energy in the 
expansion to achieve convergence. When $\Omega = 2.5N$, the 
difference is already not so significant; and when $\Omega=4.5N$, 
we can clearly see that the contribution of $\Sigma_2(\tilde{\omega})$ is 
negligible. Therefore, our results should be accurate enough as 
long as $\Omega>2.5N$. 
For unpolarized nuclei, $\Omega=2.5N$ 
corresponds roughly to an external field of 5 Tesla's for GaAs 
dots. \cite{Paget_PRB_77}

We have checked the sum rule for the spectral function, 
$\int_{-\infty}^{\infty} \rho(\tilde{\omega})d\tilde{\omega}=1$, 
in all the numerical calculations. We find that the 
sum rule is always satisfied within $10^{-3}$.
% for the calculations 
%using only $\Sigma_1({\tilde{\omega}})$,
%and $10^{-3}$ for those with both $\Sigma_1(\tilde{\omega})$ and 
%$\Sigma_2(\tilde{\omega})$. This check also ensures that our 
%evaluations of the self-energy should be correct. 
%because the 
%real and imaginary part of the self-energy have to satisfy the 
%Kramers-Kronig relation, 
%otherwise the sum rule won't be satisfied. 
Fig. \ref{Spectrum} also shows that the original delta function 
$\delta\left[\omega-\Omega+\text{O}(1)\right]$ without decoherence
has been broadened to a continuous spectrum after the 
electron-mediated flip-flop of nuclear spins are explicitly 
considered. The sum rule from the continuous spectrum clearly indicates
that there is no contribution from a delta-function to the spectral 
function, so that the decay of $G_{\perp}(t)$ should be complete.
 
\begin{figure}[t]
\begin{center}
\epsfig{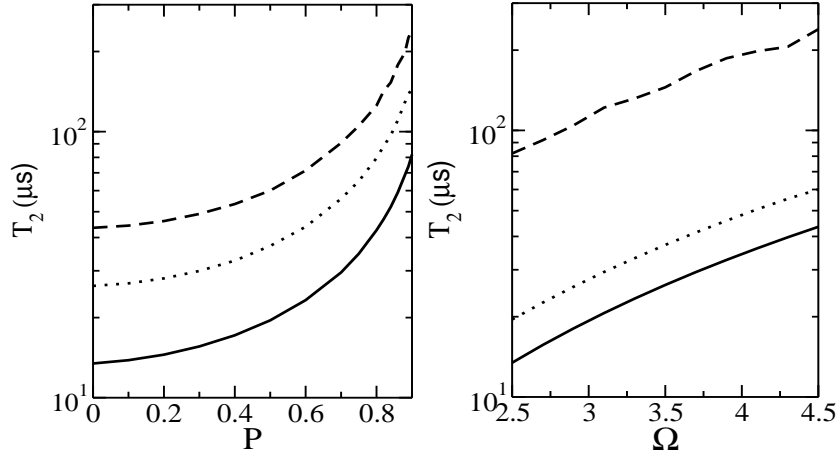}
\caption{The electron spin $T_2$ time in logarithm scale as a function
of nuclear spin polarization $P$ (left panel) and the effective
magnetic field $\Omega$ (right panel). $T_2$ is estimated with
the spectral function by finding the half-width of the spectral
peak. It is easy to see that the coherence time can greatly 
enhanced by increasing the nuclear polarization to 80\%. We have 
assumed $A=92 ~\mu$eV in these calculations.}
\label{T2}
\end{center}
\end{figure}

Once the spectral function is obtained, the electron spin dephasing 
time $T_2$ can be
estimated as the half-width of the spectrum at $\tilde{\omega}=0$ peak, i.e., 
$T_2 = 1/\Delta \tilde{\omega}$. The result is shown in Fig. \ref{T2}, where
the dephasing time is plotted as functions of the nuclear polarization $P$
and the effective magnetic field $\Omega$. 
In all the numerical calculations, we have assumed the summed 
hyperfine coupling constant $A=\sum_k A_k=92 ~\mu eV$.
Fig. \ref{T2} indicates that to increase $T_2$,
polarizing the nuclear spins is more effective than increasing 
the external magnetic field. 
In essence, increasing $P$ leads to a reduction to in the phase space
for nuclear spin flip-flop, while increasing $\Omega$ leads to increased
energy for the intermediate state and thus reduced cross-section for the
higher-order process.
The $T_2$ time can be enhanced by almost one
order of magnitude by increasing $P$ from 0 to $90\%$, while 
applying higher magnetic fields 
extends the coherence time by a few times. For unpolarized nuclei 
with $\Omega=2.5N$, we 
find $T_2 \approx 10$ microsecond, which is similar to the 
decoherence time caused by the nuclear dipole-dipole interaction. 
This comparison needs to be kept in perspective, however, 
since our numerical results are calculated for an effectively
two-dimensional QD with Gaussian type wave function, while the previous
results for dipolar coupling are obtained for a 3-dimensional QD.
\cite{Roger_PRB_03} 
It would be interesting 
to explore how  dimensionality changes the $T_2$ times quantitatively.
In addition, our results
%Another aspect that
%could change the results in Figure \ref{T2} is that we have been 
are obtained by assuming nuclear
spin $I=\frac{1}{2}$ throughout this paper, though all the isotopes
of Ga and As nuclei have spin $I=\frac{3}{2}$. 
Exploration of the dimension and $I$ dependence of $T_2$ 
would be interesting, but goes beyond the scope of the present paper.
%In all the numerical calculations, we have assumed the summed 
%hyperfine coupling constant $A=\sum_k A_k=92 ~\mu eV$.

Obviously the $T_2$ time is proportional to the unit $N/A$. Therefore a 
larger quantum dot [with a larger number of nuclear spins ($N$)] has a
longer $T_2$ time. In the bulk limit the coherence time becomes
infinitely large because the hyperfine coupling is homogeneous so that
there is no fluctuation of nuclear field. In the other limit of a
smaller quantum dot the profile of the electron wavefunction becomes
sharper because of the strong confinement. One thus also
anticipate increased $T_2$ time since the effective nuclear flip-flop
is more difficult to realize energetically just like in the case of the
dipolar coupling. \cite{Roger_PRB_03} These arguments indicate that
there should be a certain point regarding the QD size where the
decoherence effect is the most serious. However, in our present
calculations we would not be able to reach the latter limit because we 
cannot discuss the influence of the distance among nuclear spins on the
electron-mediated interaction when converting the sum over nuclear sites
into integrals.

\begin{figure}[t]
\begin{center}
\epsfig{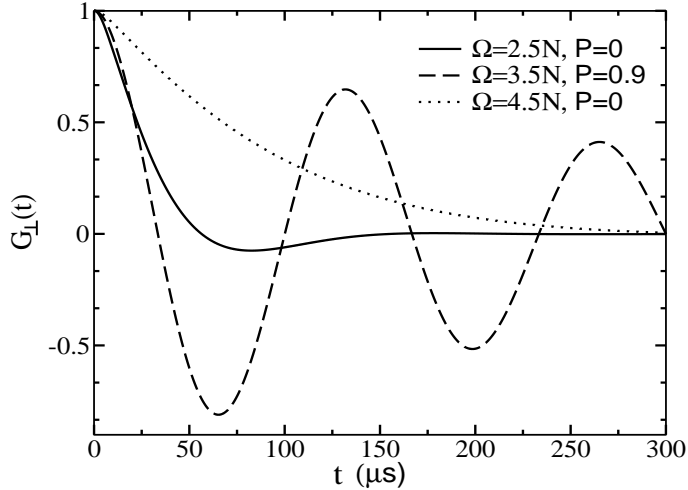}
\caption{The real part of the Green's function $G_{\perp}$ for
different regime of $T_2$. $G_{\perp}$ is evaluated by performing the
inverse Fourier transform with Eq. (\ref{inv_Four}). We have only plotted the envelope
function of the correlation function. The actual evolution should be
modulated by a fast oscillation with frequency $\Omega$.}
\label{Evolution}
\end{center}
\end{figure}

The real time dynamics of the transverse electron spin Green's function 
$G_{\perp}(t)$ can be
obtained by performing the inverse Fourier transform using 
Eq. (\ref{inv_Four}).
In Fig. \ref{Evolution} we show the evolution of the envelope 
of $G_{\perp}(t)$ (the maximum values during each cycle)
for various $P$ and $\Omega$ combinations. The actual evolution
of $G_{\perp}(t)$ oscillates coherently with frequency close to $\Omega$.
In the relative small-magnetic-field ($\leq$10T) and low-polarization limit, 
the amplitude of $G_{\perp}(t)$ decays rapidly and completely. On the other
hand, the coherence of the electron spin can be maintained for a much longer
period if $P=0.9$ even without a very large $\Omega$. We should mention 
that the 90\% nuclear polarization could be realizable in future 
experiments using dynamical nuclear polarization by polarized
electronic transport \cite{Ono_PRL_04, Deng_PRB_05} or 
circularly polarized photons \cite{Gammon_PRL_96, Imam_PRL_03, Lai_PRL_06}
through hyperfine interaction.
The current experimental realizable nuclear polarization in QDs
remains less than 50\%. \cite{Gammon_PRL_96} Our 
approximation should still be accurate with such a large 
nuclear polarization,
because the numbers of up and down nuclear spins are still 
in the same order 
of magnitude at $P=0.9$. If we go to the extremely polarized limit, the
decoherence effect should become negligible as in the case of fully polarized 
nuclei. For the curve with $P=0.9$, we also observe that 
there is a clear revival for $G_{\perp}(t)$ 
even after 100 microseconds.

With our approach we can identify the long time asymptotic behavior 
of $G_{\perp}(t)$. 
Again, the contour shown in Fig. \ref{loop_1} can be used to calculate 
the asymptotic integrals of Eq. (\ref{inv_Four}) as $t\rightarrow \infty$ 
with the method of 
steepest descent. For simplicity we will only consider 
the effect of $\Sigma_1(\tilde{\omega})$. To evaluate these integrals we 
need to find the asymptotic form of the spectral function. The calculation 
is similar to what we have done for the low frequency solution. On 
the contours of $C_2$ and $C_3$ in Fig. \ref{loop_1}, we define 
$\tilde{\omega}=0^{\mp}-is$. On the contours of $C_1$ and $C_4$, we replace
$\tilde{\omega}$ by $\mp \frac{1}{2}-is$. Here $s$ is a new real variable.
The long time ($t\rightarrow \infty$) behavior is determined by the 
asymptotic form of the spectral function as $s \rightarrow 0^+$. 
On $C_2$ and $C_3$, we find
\be
\rho(0^{\mp} - is) = \frac{24\Omega^2}{\pi^2(1-P^2)N^2} (1\mp3is) 
+ \text{O}(s^3).
\ee
Performing the integral along the contours $C_2$ and $C_3$ as 
$t\rightarrow \infty$, we find
\be
G_{\perp}(t) \propto \frac{144\Omega^2}{\pi^2(1-P^2)N^2} \frac{1}{t^2}.
\label{long_time_G}
\ee
On the contours $C_1$ and $C_4$, 
$\rho(\mp \frac{1}{2}-is) \propto s^2$.
Therefore, the integral along $C_1$ and $C_4$ has a contribution of 
$\frac{1}{t^3}$ which is negligible compared to $\frac{1}{t^2}$ at 
long time limit.
This non-exponential long time decay behavior 
is slower than that of the low-energy solution with strong magnetic 
field, where the form of $\frac{1}{t}$ is found. 
The $\frac{1}{t^2}$ behavior here is obtained by only considering 
$\Sigma_1(\tilde{\omega})$. The inclusion of $\Sigma_2(\tilde{\omega})$
is too complicated. However, we would expect power-law decay instead 
of the $\frac{1}{\text{ln}t}$ decay which appears in the solution 
for the low-energy solution in the absence of external magnetic field.
%energy solution in the absence of external magnetic field.
%Nevertheless, it is expected that 
%we should get a 
%power-law decay instead of the $\frac{1}{\text{ln}t}$ decay of the low 
%energy solution in the absence of external magnetic field. 
Notice that the expression in Eq. (\ref{long_time_G}) diverges as the nuclear 
polarization goes toward 1. However, this is insignificant since
our discussions in this subsection apply to partially and unpolarized 
nuclei so that $P$ is always less than 1. Instead,
the low energy solution has the right limit when $P=1$. This is
because both the exact solution of fully polarized nuclei and the
low energy solution include only the real electron-nuclei flip-flop,
while the high energy solution found in this section is due completely
to the higher-order processes, which are only significant when $P<1$.

\section{Discussion}
The calculations given in this paper can be generalized 
in several ways. Firstly, although we have 
been dealing with nuclei of spin $\frac{1}{2}$ exclusively throughout 
this paper, our study can be easily extended to 
nuclei with higher spin values. The qualitative behavior of the 
spectral function and the long time 
decay we have discovered here should not be modified 
in any significant way by considering 
nuclei with larger spins. Secondly, we have assumed a relatively 
simple form of the hyperfine coupling constant
as a function of the position of the nucleus 
at $\mathbf{r}_k$, i.e., $A_k=e^{-\frac{k}{N}}$, to simplify the algebra. 
This form of $A_k$ corresponds to a two-dimensional quantum dot 
with Gaussian electron wavefunction. However, our 
derivations and approximations are not limited by this choice of $A_k$ and
are applicable to any form of quantum dot with arbitrary electronic 
wavefunction. In short,
all our results before converting the summations into integrals 
are correct for a general electron wavefunction.
Generalization to a 3-dimensional quantum 
dot is more complicated mathematically because one would 
encounter integrations that cannot be performed analytically when 
converting the sums over nuclear lattice sites to integrals. 

A more challenging and interesting generalization of the 
present work is to consider smaller effective magnetic fields, 
for example with $\Omega \sim \text{O}(\sqrt{N})$. It is well-known 
that the thermal statistical
fluctuating nuclear magnetic field has this order of 
magnitude. \cite{Book_OO} In such
a case, our large $N$ expansion can still be applied, because 
$\sqrt{N}$ is still much larger than $A_k$. The Zeeman energy mismatch
determines that the direct electron-nuclei relaxation effect is 
still negligible.
However, the large field
expansion we have developed in this paper loses its power in this 
situation since the expansion 
parameter $\frac{N}{\Omega} \gg 1$. Basically one has to consider 
more higher-order correlation functions with more pairs of 
flip-flopping nuclear spins. We expect this to be a very
difficult task for analytical calculations. 
%However, our treatment
%can be applied to study the spin dynamics of spin-boson model, where 
%a single spin interacts with a reservoir of bosons at thermal 
%equilibrium

We have studied the Non-Markovian dynamics of electron spin with 
the nuclear spin reservoir in an initial product state. 
It has recently been found that the electron spin dynamics shows
different behaviors for randomly correlated initial nuclear spin states 
or ensemble averaged initial states. \cite{Schli_PRB_02}  
Our analytical results can be directly used to investigate the 
electron spin decoherence by summing over the product states in 
the ensemble. Our treatment can also be applied to study the 
Non-Markovian spin dynamics of spin-boson model, 
where a single spin interacts with a reservoir of bosons at 
thermal equilibrium in the initial state. It would also be 
interesting to extend the approach presented in this paper to 
study the dephasing time of two-electron spin 
states in double QD, where both the $T_2^*$ and $T_2$ has been
measured. \cite{Petta_Science_05}

\section{Conclusions}

To summarize, we have obtained the {\it decoherence} of a single 
electron spin confined in a
quantum dot with large effective magnetic fields (including both 
external field and nuclear field)
by computing the Green's function $G_{\perp}(t)$ for the electron spin
using the equation-of-motion method. 
We have solved this problem for three types of initial nuclear spin states, 
the fully polarized case,
the almost fully polarized case with one nuclear spin flipped, 
and the more general partially polarized and
unpolarized cases. We have found that the spectral functions have
quite different properties at the low frequency 
($\omega \sim A/N$) and the high frequency 
region ($\omega \sim A$). By comparing the exact solution of fully polarized
nuclei and the solution of almost fully polarized cases, we demonstrate
that the virtual nuclear spin flip-flop leads to a new
feature in the high energy limit. Our studies of electron spin
decoherence in the low energy region 
recover previous results, which did not include these higher-order
processes. Namely, the decay amplitude is of the order 
$\text{O}\left( \frac{1}{N}\right)$. More importantly,
we have also obtained the solution of partially polarized and unpolarized 
nuclei by considering the indirect nuclear spin flip-flop explicitly, helped
with a large field expansion method. We find that the electron spin $T_2$ 
time is very sensitive to nuclear polarization when $P>0.6$, and the 
electron spin coherence time can be enhanced by 10 times if the nuclear 
polarization is increased to 90\%. We also find that this decoherence is 
complete, and the long time asymptotic behavior of the Green's function
representing the transverse electron spin dynamics is $\frac{1}{t^2}$.

\acknowledgments
We acknowledge financial support by NSA, LPS, and ARO.

\appendix
\section{Exact Solution of Fully Polarized Nuclei in 
Jordan-Wigner Representation}
\label{App_JW}
For spin one-half nuclei, we can introduce the Jordan-Wigner 
representation: \cite{Book_QP1D} 
\be\label{JW}
S^{+} &=& d, ~S^{-} = d^{+}, ~S^{z} = \frac{1}{2} - n_d\nonumber \\
I_k^{+} &=& e^{-i\pi n_d} a_k, ~I_k^{-} = a_k^{+}e^{i\pi n_d}, 
~I_k^{z} = \frac{1}{2} - n_k,
\ee
where $n_d=d^{+}d$ and $n_k=a^{+}_ka_k$. All operators obey the standard 
fermion anti-commutation relations: 
$\{d,d^{+}\} = 1$; $\{a_k,a_k^{+}\}=1$. This representation preserves
all spin commutation relations except for those of two nuclear spin 
operators at two different lattice sites. However these commutators 
are not necessary for fully polarized nuclei, because  
there should be only one flipped spin at any moment 
during the evolution. In terms of correlation functions, this means we 
won't encounter any correlation function containing two nuclear
operators so that there is no problem of ordering different nuclear 
spin operators using the fermion operators. 

Transforming the original Hamiltonian of 
Eq. (\ref{ham_spin}) into the new representation, we arrive at
\begin{eqnarray}
H_{\text{JW}} &=& -\sum_k \frac{A_k}{2}n_k 
- \left(\omega_0 + \frac{A}{2} \right)n_d
%+ \sum_k A_kn_kn_d \nonumber \\
\nonumber \\
&+&\sum_k \frac{A_k}{2}\left(a_k^+d + d^+a_k \right ),
\label{H_JW}
\end{eqnarray}
where $A=\sum_k A_k$. We have ignored a constant term in the 
derivation. The quartic interaction term $\sum_kA_kn_kn_d$ in 
$H_{\text{JW}}$ is dropped because $\sum_kA_kn_k$ and $n_d$ 
cannot both be 1 (down state). The new 
Hamiltonian $H_{\text{JW}}$ is in the form of the non-interacting Anderson 
impurity model, \cite{Anderson_PR_61} describing a 
localized state (electron spins) interacting with semi-continuous 
states represented by the nuclear spins. This is the key feature 
that we focus in this paper: a single electronic spin interacts with 
many nuclear spins with different strengths. The non-interacting
Anderson impurity model can be solved exactly. Specifically, 
the exact solution of $H_{\text{JW}}$ for the 
Green's function $\langle\langle d;d^+\rangle\rangle_{\omega}$ 
is \cite{Book_Mahan}
\be
\langle\langle d;d^+\rangle\rangle_{\omega}
=\frac{1}{\omega + \omega_0+\frac{A}{2}
-\frac{1}{4}\sum_{k}\frac{A_k^2}{\omega+\frac{A_k}{2}}}.
\ee
The spectral function 
($-\text{Im}\langle\langle d;d^+\rangle\rangle_{\omega}/\pi$)
that represents the overlapping of the localized state 
with the exact eigenstate is the same as what we have found in 
Section III.

\section{EOMs for the Polarized Nuclear Reservoir with One-Flipped Nuclear Spin}
\label{App_EOM_One_Flipped}
The EOMs are generated by computing the commutators in Eq. (\ref{eom}). 
%The calculations are tedious but straightforward. 
The highest order 
correlation functions that survive are those which involve two 
spin-lowering operators, either $S^-$ or $I^-_k$. All higher order 
functions with more spins flipped to down direction vanish 
because the total angular ($\sum_k A_kI_k^z+S^z$) 
momentum along $z$ direction (of the external field)  
is a constant of motion. Explicitly, the equations are:
\begin{widetext}
\begin{eqnarray}
&&\left [ \omega - \Omega - \Sigma_0(\omega) + A_k + 
\frac{A_k^3}{4(\omega^2-A_k^2/4)}\right ]
\la\la n_kS^-;S^+\ra\ra_{\omega} = \la\Psi_0|n_k|\Psi_0 \ra \nonumber \\
&&-\frac{1}{4}\sum_{k'(k)}\frac{A_kA_{k'}V_{kk'}(\omega)}{\omega+A_k/2}
-\frac{1}{4}\sum_{k'(k)}\frac{A_kA_{k'}V_{k'k}(\omega)}{\omega-A_{k'}/2}, \\
\label{EOM_One_1}
%%%%%%%%%%%%%%%%%%%%%%%%%%%%%%%%%%%%%%%%%%%%%%%%%%%%%
&&\left (\omega - \frac{A_{k'}}{2} \right)
\la\la n_kI_{k'}^-;S^+\ra\ra_{\omega} = 
-\frac{A_{k'}}{2}\la\la n_kS^-;S^+ \ra\ra_{\omega} 
+\frac{A_k}{2}V_{kk'}(\omega), \\
\label{EOM_One_2}
%%%%%%%%%%%%%%%%%%%%%%%%%%%%%%%%%%%%%%%%%%%%%%%%%%%%%
&&\left ( \omega + \Omega - \frac{A_k+A_{k'}}{2} \right ) 
\la\la I_k^-S^+I_{k'}^-;S^+\ra\ra_{\omega} =
\frac{A_k}{2}\la\la n_kI_{k'}^-;S^+\ra\ra_{\omega}
+\frac{A_{k'}}{2}\la\la I_{k}^-n_{k'};S^+ \ra\ra_{\omega} \nonumber \\
&&-\frac{A_k}{2}\la\la I_{k'}^
-\left(\frac{1}{2}-S^z\right);S^+\ra\ra_{\omega}
%+\frac{A_{k'}}{2}\la\la I_{k}^-n_{k'};S^+ \ra\ra_{\omega}
 -\frac{A_{k'}}{2}\la\la I_k^-\left(\frac{1}{2}
-S^z\right);S^+\ra\ra_{\omega} \nonumber \\
&&+\sum_{k''(k,k')}\frac{A_{k''}}{2}\la\la
 I_k^-I_{k''}^+I_{k'}^-;S^+\ra\ra_{\omega}, \\
\label{EOM_One_3}
%%%%%%%%%%%%%%%%%%%%%%%%%%%%%%%%%%%%%%%%%%%%%%%%%%%%%%
&&\left( \omega-\Omega+\frac{A_k+A_{k'}}{2}\right)
\la\la I_k^-I_{k'}^+S^-;S^+\ra\ra_{\omega} = 
\frac{A_{k'}}{2}\la\la I_k^-\left( \frac{1}{2}
-S^z\right);S^+\ra\ra_{\omega} \nonumber \\
&&-\frac{A_{k'}}{2}\la\la I_k^-n_{k'};S^+\ra\ra_{\omega} 
-\sum_{k''(k,k')}\frac{A_{k''}}{2}\la\la
 I_k^-I_{k'}^+I_{k''}^+;S^+\ra\ra_{\omega}, \\
\label{EOM_One_4}
%%%%%%%%%%%%%%%%%%%%%%%%%%%%%%%%%%%%%%%%%%%%%%%%%%%%%%%
&&\left( \omega - \frac{A_k+A_{k''}-A_{k'}}{2}\right)
\la\la I_k^-I_{k'}^+I_{k''}^-;S^+\ra\ra_{\omega} = 
-\frac{A_k}{2}\la\la I_{k''}^-I_{k'}^+S^-;S^+\ra\ra_{\omega}
\nonumber \\
&&+\frac{A_{k'}}{2}\la\la I_k^-S^+I_{k''}^-;S^+ \ra\ra_{\omega} 
  -\frac{A_{k''}}{2}\la\la I_k^-I_{k'}^+S^-;S^+\ra\ra_{\omega}.
\label{EOM_One_5}
\end{eqnarray}
\end{widetext}
Together with Eqs. (\ref{Master_Eq}), these equations form a 
closed set. The full solution of these equations is 
mathematically intractable because of the number 
of equations involved.
In Section III, we find approximate solutions in the low 
frequency ($\omega \sim 1$) and 
high frequency ($\omega \sim \Omega$) regions 
using $\frac{1}{N}$ expansion. 
%See our complete discussions of one-flipped nuclear 
%spin configuration in Sec. III.

\section{Evaluation of $\sigma_n(\omega)$}
\label{App_self_low}
In this appendix the real and imaginary parts of the 
terms $\sigma_n(\omega)$ [defined in Eq. (\ref{self_low})] appearing in 
the low energy solution of partially polarized and 
unpolarized nuclei in Section III are 
calculated. The spectral function of the 
electron spin correlation function calculated from
$\sigma_{n}(\omega)$ can be used to obtain the 
renormalized spin precession frequency and decay of 
electron spin coherence.
There are three steps in these calculations. 
First we perform analytical 
continuation by replacing $\omega$ with $\omega + i0^+$ to obtain 
the retarded expressions. 
We then use the relation 
$\frac{1}{x+i0^+}= \text{P}\frac{1}{x}-i\pi\delta(x)$ 
to separate the principle values and the imaginary parts. 
Finally we convert the summation over the nuclear sites
into an integral, $\sum_k \rightarrow \int_0^{\infty} dk$. 
The validity and efficiency of this conversion have been 
discussed before. \cite{Coish_PRB_04}  

Recall that the summation $\sigma_2(\omega)$ is
\be
\sigma_2(\omega) = \sum_k \frac{A_k}{\omega - \frac{A_k}{2}}
-\sum_k \frac{A_k}{\omega+\frac{A_k}{2}}.
\ee
Using the procedures described above we find that the real 
and imaginary parts of 
$\sigma_2(\omega)$ are
\be\label{S1_real}
\text{Re}~\sigma_2(\omega) = -2N\left[~\text{ln}\left|1+\frac{1}{2\omega}\right|
 + \text{ln}\left|1-\frac{1}{2\omega}\right|~\right],
\ee
and
\be
\text{Im}~\sigma_2(\omega) = \left \{
  \begin{array}{ll}
    -2N\pi & ~ 0<\omega<\frac{1}{2} \\
    2N\pi& ~-\frac{1}{2}<\omega < 0.
   \end{array} \right.
\ee
$\sigma_3(\omega)$ and $\sigma_4(\omega)$ can be evaluated in a
similar manner. We obtain
\be
\text{Re}~\sigma_3(\omega) = -4N+4N\omega\text{ln}
\left|\frac{2\omega+1}{2\omega-1}\right|,
\ee
\be
\text{Im}~\sigma_3(\omega) = -4N\pi\omega, ~-\frac{1}{2}<\omega<\frac{1}{2},
\ee
\be
\text{Re}~\sigma_4(\omega) = -2N-8N\omega^2\left[~\text{ln}\left|1+\frac{1}{2\omega}\right|
 + \text{ln}\left|1-\frac{1}{2\omega}\right|~\right],
\ee
and
\be
\text{Im}~\sigma_4(\omega) = \left \{
  \begin{array}{ll}
    -8N\pi\omega^2 & ~ 0<\omega<\frac{1}{2} \\
     8N\pi\omega^2 & ~ -\frac{1}{2}<\omega < 0,
   \end{array} \right.
\ee
Both $\omega =\pm \frac{1}{2}$ and $\omega=0$ are branch 
points for the self-energy.
The two branch cuts ([-1/2,0] and [0,1/2]) come from 
different dynamical fields of the electron felt by the nucleus, i.e., 
$\frac{A_k}{4}$ when
$S^z=\frac{1}{2}$, and $-\frac{A_k}{4}$ when 
$S^z=-\frac{1}{2}$. 
%We have employed the Fermion operators defined in appendix A. 
In contrast only $\omega=-\frac{1}{2}$ appears as a branch point in the 
fully polarized case because $S^z=1/2~(n_d=0)$ makes no contribution
to the Hamiltonian in \ref{H_JW}.

\section{Evaluation of Self-energy Terms, 
$\Sigma_1(\tilde{\omega})$, $\Sigma_2(\tilde{\omega})$, 
and $\Sigma_3(\tilde{\omega})$}
\label{App_self_high}
Using the definitions of the self-energy terms given in Eqs. (\ref{Self_1}),
(\ref{Self_2}) and (\ref{Self_3}), and converting the summations into integrals,
we can obtain analytical expressions of the real and imaginary parts of 
these self-energy terms. The same procedures as we have described in 
Section \ref{App_self_low}
are used in the following calculations. We find
\be
\Sigma_1({\tilde{\omega}}) = \sum_{k,k'}
\frac{A_k^2A_{k'}^2}{\tilde{\omega}+\frac{A_k-A_{k'}}{2}}
=N^2\int_0^1 dx\int_0^1 dy \frac{xy}{\tilde{\omega}+\frac{x-y}{2}},
\ee
where we have written $A_k$ and $A_{k'}$ as the integral variables $x$ and $y$.
The two-dimensional integral can be calculated, so that
\begin{eqnarray}
\text{Re}~\Sigma_1(\tilde{\omega}) &=&
-\frac{2N^2}{3} \left[ \tilde{\omega}
                  + 4\tilde{\omega}^3\text{ln}
                   \left|1-\frac{1}{4\tilde{\omega}^2}\right| 
                \right. \nonumber \\
                &-& \left. 3\tilde{\omega}\text{ln}
                \left| 1-\frac{1}{4\tilde{\omega}^2}\right|
                + \text{ln}\left|
                \frac{2\tilde{\omega}-1}{2\tilde{\omega}+1}\right|
                \right],
\end{eqnarray}
and
\be
\text{Im}~\Sigma_1(\tilde{\omega}) = 
-\frac{2N^2}{3}\left[ 4|\tilde{\omega}|^3-3|\tilde{\omega}|+1\right].
\ee
Again, the imaginary part of $\Sigma_1(\tilde{\omega})$ is nonvanishing 
only when $-\frac{1}{2}<\tilde{\omega}<\frac{1}{2}$. Similarly,
\begin{eqnarray}
\Sigma_2({\tilde{\omega}}) = N^3\int_0^1 dx \int_0^1 dy \int_0^1 dz
\frac{xyz}{(\tilde{\omega}+\frac{x-y}{2})
           (\tilde{\omega}+\frac{x-z}{2})}.
\end{eqnarray}
\begin{eqnarray}
\text{Re}\Sigma_2(\tilde{\omega}) = 4N^3\int_0^1 ds ~s
\left[ 1 + (2\omega+s)\text{ln}\left| \frac{2\tilde{\omega}+s-1}
       {2\tilde{\omega}+s}\right| \right]
\label{Re_S2}
\end{eqnarray}

\be
\text{Im}~\Sigma_2(\tilde{\omega}) = \left \{
  \begin{array}{ll}
    \frac{1}{2}-\text{ln}|1-2\tilde{\omega}| -2\tilde{\omega}
     + \frac{8\tilde{\omega}}{3}\text{ln}|1-2\tilde{\omega}| 
     + \frac{10\tilde{\omega}^2}{3} \nonumber \\
     -\frac{8\tilde{\omega}^3}{3} 
     -\frac{16\tilde{\omega}^4}{3}
      \text{ln}\left|\frac{1}{2\tilde{\omega}}-1 \right|, 
     ~0<\tilde{\omega}<\frac{1}{2} \nonumber \\
     \\
     \frac{1}{2}-2\tilde{\omega}-8\tilde{\omega}
     \text{ln}|1-2\tilde{\omega}|+\frac{16\tilde{\omega}^2}{3}
     \text{ln}|1+2\tilde{\omega}| \nonumber \\
     +8\tilde{\omega}^2\text{ln}\left|\frac{1}{2\tilde{\omega}}+1\right| 
     -\frac{14\tilde{\omega}^2}{3} + \frac{8\tilde{\omega}^3}{3} \\
     -\frac{16\tilde{\omega}^4}{3}
     \text{ln}\left|\frac{1}{2\tilde{\omega}}+1 \right|,
     ~-\frac{1}{2}<\tilde{\omega} < 0,
   \end{array} \right.
\ee
Repeating the calculation for $\Sigma_3(\tilde{\omega})$, we find that
$\Sigma_3(\tilde{\omega}) = \Sigma_2(\tilde{\omega})$. The integral in 
Eq. (\ref{Re_S2}) can be computed numerically. However,
the calculation is non-trivial because of the singularities appearing in
the expression. Alternatively, the integral can be done analytically 
using Maple. The result, which is too complicated to be presented here, 
is a sum of terms that include the dilog functions defined as
$f_{\text{dilog}}(x)=\int_0^x\text{ln}t/(1-t)dt$. \cite{Handbook} 
Numerical calculation of the real part of $\Sigma_2(\tilde{\omega})$ 
using the analytical expression obtained with Maple is very accurate
when we check the sum rule of the spectrum function 
numerically (see discussion in Section III).

\end{document}